\documentclass[12pt,a4paper]{JHEP3}

\usepackage{amscd,amsmath,amssymb,amsfonts,xspace,mathtools}
  \usepackage{xcolor}
  \usepackage{tikz-cd}
  \let\normalcolor\relax      
\usepackage{latexsym}  
 \usepackage{graphicx}

\usepackage{graphicx, subfig, float}

\usepackage{epsfig,amsfonts,amssymb}
\usepackage{framed}
\newcommand{\refb}[1]{(\ref{#1})}
\newcommand{\Tr}{\mathop{\mathrm{Tr}}}

\newcommand{\nn}{\ensuremath{\nonumber{}}}

\title{Black holes and the loss landscape in machine learning}

\author{Pranav Kumar$^1$, Taniya Mandal$^2$, Swapnamay Mondal$^3$ 
\\
$^1$ {\it Capgemini Engineering R \& D, Kadubisanahalli, Bengaluru, 560103, India} \\
$^2$ {\it Department of Physics, Indian Institute of Technology Kanpur, Kalyanpur, Kanpur 208016, India}
\\
$^3$ {\it Dublin Institute for Advanced Studies, 10 Burlington Road, Dublin, Ireland}
\\
\vspace*{2mm}\\
{\tt e-mail: pranavkr29@gmail.com, taniyam@iitk.ac.in, swapno@maths.tcd.ie }
\vspace*{-3mm}
}  
 
\abstract{
Understanding the loss landscape is an important problem in machine learning. 
One key feature  of the loss function, common to many neural network architectures, is the presence of exponentially many low lying local minima. 
Physical systems with similar energy landscapes may provide useful insights.
In this work, we point out that black holes naturally give rise to such landscapes, owing to the existence of black hole entropy. 
For definiteness, we consider 1/8 BPS black holes in $\mathcal{N}=8$ string theory. These provide an infinite family of potential landscapes arising in the microscopic descriptions of corresponding black holes. The counting of minima amounts to black hole microstate counting. Moreover, the exact numbers of the minima for these landscapes are a priori known from dualities in string  theory. 
Some of the minima are connected by  paths of low loss values, resembling mode connectivity.
We  estimate the number of runs needed to find all the solutions. Initial explorations suggest that Stochastic Gradient Descent can find a significant fraction of the minima.
}

\begin{document}
 
\maketitle

\pagebreak
\section{Introduction and summary}
Machine learning has achieved tremendous success in various applications of practical interest \cite{NIPS2012_c399862d, 5740583, Manning}, as well as in more academic endeavours involving large datasets. These include, but are not limited to, string theory landscape \cite{He:2017aed,He:2017set,Ruehle:2017mzq,Carifio:2017bov,Mutter:2018sra,He:2021eiu}, 
Calabi Yau manifolds \cite{Bull:2018uow, Bull:2019cij,Jejjala:2020wcc,Berglund:2022gvm,Erbin:2020srm,Erbin:2020tks,He:2020lbz,Anderson:2020hux,Berman:2021mcw,Erbin:2021hmx},
knot invariants \cite{Craven:2020bdz,Craven:2021ckk,Craven:2022cxe, Gukov:2020qaj}, 
holography \cite{Hashimoto:2018ftp,Hashimoto:2018bnb,Tan:2019czc,Akutagawa:2020yeo, Yan:2020wcd}, conformal field theory \cite{Chen:2020dxg, Basu:2022qaf, Kuo:2021lvu, Kantor:2021kbx,Kantor:2021jpz,Kantor:2022epi}, Lie groups \cite{Chen:2020jjw,Lal:2022otf}. Connections between deep learning and quantum field theory have been explored in \cite{Koch:2019fxy,Halverson:2020trp,Maiti:2021fpy,Halverson:2021aot, Erbin:2021kqf,Grosvenor:2021eol,Erbin:2022lls,Banta:2023kqe}.
 Also see \cite{CaboBizet:2020cse,Hashimoto:2019bih,Betzler:2020rfg,Krippendorf:2020gny,Bao:2020nbi,Ruehle:2020jrk,He:2023csq, bedolla2020machine,samarakoon2021machine,carrasquilla2017machine,decelle2022introduction,RevModPhys.91.045002} for other applications of machine learning in Physics.  

Most machine learning tasks amount to choosing an optimal Neural Network (NN). This  is usually done by first defining a loss function, which depends on various parameters of the NN (and number in millions in realistic applications), then one finds a suitable minima of the loss function, often by Stochastic Gradient Descent (SGD) \cite{EonBottou1998OnlineLA}. 
Thus gaining an understanding of the loss function landscape is crucial for machine learning.

The loss function, also referred to as the cost function,  is in general a non-convex function. Finding a global minimum of a general non-convex function is hard \cite{10.1162/089976602760408035, livni2014computational, shalevshwartz2017failures}, in fact NP-hard \cite{Murty1987SomeNP}, \cite{10.5555/2969735.2969792}.
Nevertheless in practice optimizing a Deep Neural Network (DNN) is usually possible. This has led many to believe that after all, for DNN-s the loss function does not have spurious local minima \cite{freeman2017topology, hoffer2018train, soudry2016bad}.

This is indeed the case for shallow linear NN-s, which have only saddles and no local minima \cite{BALDI198953}. For  deep linear networks, all local minima of the  squared loss function are global, with other critical points being saddles \cite{kawaguchi2016deep}. It has been further argued  that the same holds true for non-linear networks, under certain assumptions \cite{kawaguchi2016deep}. 
For a fully connected network with squared loss and analytic activation function, all minima have been argued to be global \cite{nguyen2017loss}.  Similar results were previously shown in \cite{107014, fra, 410380}.

Although optimization of deep linear models exhibits some similarities with optimization of deep non-linear  models \cite{saxe2014exact},  ``local implies global'' picture seems not to carry over to non-linear DNN-s.
E.g. spurious minima have been shown to exist for shallow Rectified Linear Unit (ReLU) networks \cite{safran2018spurious}. It  has been suggested that absence of spurious local minima might be a feature specific to deep linear networks and might be destroyed as soon as non-linearity is introduced  \cite{yun2019small, zou2018stochastic, swirszcz2017local}.  For Convolutional Neural Networks (CNN) with  piecewise linear activation function, spurious minima have been argued to be common \cite{liu2021spurious}.

Local or global, exponentially many minima seems to be a generic feature of the loss function. 
Typically DNNs involve millions of parameters, but even for moderately sized networks, the number of local optima and saddle points already shows exponential growth in the number of parameters \cite{NIPS1995_3806734b, NIPS1996_a51fb975, choromanska2015loss, pmlr-v38-choromanska15}.
Existing evidence suggests that critical points that lie high on the loss landscape are more likely to be saddles, whereas low-lying critical points are more likely to be local minima \cite{Ramu-2005, Bray_2007, fyodorov2007replica}. It has been argued that the main obstacle in finding global minimum comes not from local minima, but from saddles \cite{dauphin2014identifying, pascanu2014saddle}.

Apart from these generalities, realistic DNN-s are too complicated to admit much analytical understanding.
Thus it is useful to study simpler models, which captures broad features  of DNN-s \cite{Goodfellow-et-al-2016}. 
Physical systems are particularly appealing in this regard, as they can often provide a physics way of thinking about NN optimization.

Connection to Physics is usually made by interpreting the loss function as some sort of energy. Glassy systems present natural candidates as their energy landscapes necessarily have many minima. Indeed, connection between NN-s and spin glasses has been known for quite some time \cite{PhysRevA.32.1007, Nakanishi_1997}. 
More recently, the loss function of a fully-connected feed-forward deep network with ReLUs  has been related to the Hamiltonian of a spin glass model \cite{choromanska2015loss, pmlr-v38-choromanska15, pmlr-v40-Choromanska15}. Critical points in these models have a layered structure \cite{auffinger2011random, Auffinger_2013}. 
Nevertheless, similarities of DNN and glassy systems are limited \cite{Baity_Jesi_2019}. Most significantly, getting stuck at local minima for long time is the quintessential glassy behaviour \cite{bouchaud1997equilibrium, cugliandolo2002dynamics, Berthier_2011}, whereas  DNN-s somehow manage to find their minima. 

Other approaches to understand the loss landscape include algebraic geometric approach \cite{mehta2018loss} and energy landscape approach \cite{C7CP01108C}. The later is known to be particularly useful in study of molecules \cite{Wales}.

Our goal in this work is rather humble. We point out that the physical potential of certain microscopic descriptions of black holes in string theory capture some key features of the loss landscape. 
The motivation for this work is to relate exponential degeneracy of loss function minima to the exponential degeneracy of quantum states in a statistical mechanical system.  This connection can be made explicit for supersymmetric black holes, where the black hole entropy implies the existence of exponentially many degenerate states, which in turn can be related to the minima of a potential through supersymmetric quantum mechanics.
Whereas connection between statistical mechanics and machine learning has been explored in the literature  \cite{doi:10.1146/annurev-statistics-032921-013738, doi:10.1146/annurev-conmatphys-031119-050745}, this particular  angle has not been  explored to the best of our knowledge.

A black hole carries an entropy $S_{BH} = \frac{A}{4 \ell_P^2}$, called Bekenstein-Hawking entropy \cite{Bekenstein1, Bekenstein2, BCH, Hawking75}. Here $A$ is the area of the event horizon of the black hole and $\ell_P= \sqrt{\frac{G \hbar}{c^3}}$ is the Planck length, G is the Newton's constant, $\hbar$ is the  Planck's constant and $c$ is velocity of light. Providing  a statistical mechanical interpretation of this entropy is a highly non-trivial challenge, supposed to be addressed by a theory of quantum gravity, such as string theory. 

In deed, $S_{BH}$ is reproduced from microscopic state counting in a wide variety of black holes in string theory \cite{Strominger:1996sh, Maldacena:1997de, Shih:2005qf}. It is done by providing a ``microscopic description'' of a black hole as a bound state of more fundamental solitonic objects in string theory, then coming up with a  low energy description  of that bound state dynamics and finally counting states carrying appropriate quantum numbers. In all known cases, this microscopic counting accounts for $S_{BH}$, when the black hole is large, which is reflected in largeness of the quantum numbers. 

Microscopic descriptions of black holes come in a wide variety in string theory. For definiteness, we consider 1/8 BPS black holes in N=8 String theory, carrying only Ramond-Ramond charges. The microscopic description of such black holes is given by bound state of D-branes, whose low energy dynamics is captured by a supersymmetric quantum mechanics with matrix and vector-like degrees of freedom \cite{Chowdhury:2014yca, Chowdhury:2015gbk}. 
The number of potential minima of this supersymmetric quantum mechanics approaches $e^{S_{BH}}$ asymptotically for large charges, i.e. for large black holes.
 Nevertheless, thanks to the dualities in string theory, the exact number of potential minima is known (without actually finding those) for arbitrary charges \cite{Shih:2005qf, Sen:2009gy}. This is a notable advantage over spin glass models for the loss landscape. E.g. testing the efficiency of an algorithm in finding different global minima is easier in current setting, as the exact number of minima is already known.

The black holes under consideration carry four kinds of electro-magnetic charges, three of which we take to be 1, and consider one parameter family of black holes labelled by the fourth charge, which we denote by $N$. For large $N$, the number of minima grows as $e^{2\pi \sqrt{N} - 2 \ln{} 4N}$. For few small $N$-s, the number of minima are given in Table \ref{indextable}. Note that all the minima are global.
\begin{table}[htb]
\begin{center}
\begin{tabular}{|c||c|c|c|c|c|c|c|c|c|c|c}
\hline
N & 1 & 2 & 3 & 4 & 5 & 6 & 7 & 8 & 9 & 10\\
\hline
number of minima  & 12 & 56 & 208 & 684 & 2032 & 5616 & 14592 & 36088 & 85500 & 195312\\
\hline
\end{tabular}
\end{center}
\caption{Number of minima of the potential landscape for small charges}\label{indextable}
\end{table}

For non-linear DNN-s, a more desirable landscape would have been one with exponentially many low lying local minima, only one of which is global. Such a landscape might be obtainable by spontaneously breaking the supersymmetry. But we shall not explore this possibility in this work.

The paper is organized as follows. In Section \ref{s2}, we give a quick introduction to black hole entropy, and explain how it arises in string theory. Readers familiar with the topic can jump to Section \ref{s3}, where we discuss the relevant potential to be minimized \refb{V_F}. In section \ref{s4}, we estimate the number of runs required to find all the minima and discuss the performance of Stochastic Gradient Descent in this regard. In section \ref{s5} we discuss future directions.

\section{A brief introduction to black hole entropy in string theory} \label{s2}
This section contains well known facts about black holes in string theory and readers familiar with the subject can skip this section.

\paragraph{A quick introduction to black hole entropy:}
Black holes are astronomical objects, formed by collapse of  dead stars. Their claim to fame is that nothing, not even light, can escape their gravitational  pull.
From a theoretical perspective, black holes are solutions to Einstein's equations with event horizon. A curious feature of such solutions is that they are entirely fixed by a handful  of parameters, such as mass, angular  momentum, electric and magnetic charges etc, and therefore leaving no scope for any microstructure and hence any entropy. However this is rather problematic, since one could violate the second law of thermodynamics, by simply throwing an entropic object into a black hole. Making use of an existing result in classical gravity that area of the event horizon  of a black hole does not go down in a physical process \cite{PhysRevLett.26.1344}, Bekenstein proposed that a black hole should be assigned an entropy proportional to the area of its event horizon \cite{Bekenstein1, Bekenstein2}. In fact it was shown in \cite{BCH} that the ``laws of black hole mechanics'' have striking resemblance with the laws of black  hole mechanics, although it was unclear whether this was a coincidence or not. In particular a black hole with non-zero temperature, is expected to emit thermal radiation, which would contradict the very  definition of a black hole.

This issue was settled by Hawking as he found that quantum mechanically black holes emit thermal radiation \cite{Hawking75}, with a temperature consistent with the findings of  \cite{BCH}. In particular, this work led to a precise formula for black hole entropy
\begin{align}
S_{BH} &= \frac{A}{4 \ell_P^2} \, , \quad  \ell_P= \sqrt{\frac{G \hbar}{c^3}} \, . \label{Sbh}
\end{align} 
This remarkable finding however led to only deeper puzzles. Now that black holes are understood to have an entropy, as  per standard statistical mechanical understanding of entropy, there must also exist $e^{S_{BH}}$ black hole  microstates. Classical gravity leaves us clueless, whereas the appearance of the Planck length $\ell_P$ in the black hole entropy formula \refb{Sbh}, suggests that answers lie in a theory of quantum gravity, such as string theory.

\paragraph{How black holes arise in string theory:}
The low energy physics of string theory is captured by supergravity theories. These supergravity theories have black hole solutions. The black holes considered in this paper, are of this type. By  the virtue of being embedded in string theory, it is often possible have an understanding of their microstates.


We will restrict to black hole solutions, that preserve some supersymmetry. Such black holes are extremal, namely ``their charge equals mass''. Here charge refers to charge (both electric and magnetic) under various U(1) gauge fields of supergravity. 
These charges are meaningful even beyond four dimensional supergravity, in particulars their carriers can be identified as membrane  like objects (called D-branes) in full string theory. This provides an alternative description of black holes as bound states of D-branes. This description, usually referred to as microscopic description, makes the microstates manifest. 

Extremal black holes carry zero temperature and hence the entropy is simply the logarithm of ground state degeneracy\footnote{The distinction between degeneracy and the index (the $14^{th}$ helicity supertrace in this case) is not too relevant for this work.}. 

\section{The potential landscape in microscopic description of black holes} \label{s3}

In most well known examples, the microscopic description of black holes is a two dimensional  Conformal Field Theory (CFT) \cite{Strominger:1996sh, Maldacena:1997de}. In such description, one looks for degeneracy of states at a given CFT level, which is captured by Cardy formula \cite{PhysRevLett.26.1344}. In this approach there is no energy landscape involved. 

Since we are looking for a potential landscape with ground states corresponding to its minima, quantum mechanical, i.e. 0+1 dimensional microscopic descriptions are better suited. For supersymmetric black holes, this quantum mechanics is supersymmetric, i.e. preserves some supercharges, which annihilate the ground states. For supersymmetric  quantum mechanics (SQM), it is known that the index $\mathcal{I} := N_B - N_F$ is  given by the Euler characteristic of the vacuum manifold, i.e. manifold on which the potential vanishes \cite{Witten_morse, Schellekens_Warner, Witten_elliptic}. Since the potential is a sum of squares in this case, vacuum manifold is also the space of minima. In general, this vacuum manifold is a continuous space and its cohomology corresponds to the space spanned by the supersymmetric ground states.
Rank  of a form decides the bosonic/fermionic nature of  the corresponding state.  

For spherically symmetric supersymmetric black holes in four space time dimensions, Zero Angular Momentum Conjecture \cite{Chowdhury:2015gbk} further implies that the black hole microstates lie in middle cohomology\footnote{The group of spatial rotations has been identified as Lefschetz SU(2) acting on the cohomology of the moduli  space \cite{Denef:2002ru, Bena:2012hf}.}, and carry zero angular momentum \cite{Sen:2009gy, Dabholkar:2010rm, Sen:2011ktd, Bringmann:2012zr}.  This in particular implies $N_F=0$, thus the degeneracy $d = N_B + N_F$ equals the index $\mathcal{I}$. 

Zero Angular Momentum Conjecture implies that the vacuum manifold is a set of points, in which case the cohomology only contains 0-forms
. 
Each discreet global minima is associated with a 0-form, i.e. a black hole microstate.
 Consequently,  the number of global minima equals the degeneracy, which is the exponential of the black hole entropy. This possibility is realized by 1/8 BPS black holes in $\mathcal{N}=8$ string theory, in a duality frame, where all charges are carried by D-branes \cite{Chowdhury:2014yca, Chowdhury:2015gbk}.

Other relevant quantum mechanical descriptions include quiver quantum mechanics \cite{Denef:2002ru}, especially scaling quivers \cite{Bena:2012hf, Chattopadhyaya:2021rdi, Beaujard:2021fsk}, which share many similarities with single centered black holes. Whereas such quivers do exhibit exponentially many states in the middle cohomology, the vacuum manifold is not a discreet set of points, hence not particularly suitable for the present purpose. 
\subsection{The system}
%
The system involves a single D2 brane stretched along the spatial directions $x^4, x^5$, a single D2 brane stretched along the spatial directions $x^6, x^7$, a single D2 brane stretched along the spatial directions $x^8, x^9$,  and N D6 branes stretched along the spatial directions $x^4, x^5,x^6, x^7,x^8, x^9$. This D-brane configuration is summarized in the Table \ref{Dconfig}.
\begin{table}[htb]
\begin{center}
\begin{tabular}{|c||c|c|c|c|c|c|c|c|c|c}
\hline
& 1 & 2 & 3 & 4 & 5 & 6 & 7 & 8 & 9 \\
\hline
D2 &  &  &  & $\checkmark$ & $\checkmark$ &  &  &  &  \\
\hline
D2 &  &  &  &  &  & $\checkmark$ & $\checkmark$ &  &  \\
\hline
D2 &  &  &  &  &  &  &  & $\checkmark$ & $\checkmark$ \\
\hline
D6 &  &  &  & $\checkmark$ & $\checkmark$ & $\checkmark$ & $\checkmark$ & $\checkmark$ & $\checkmark$ \\
\hline
\end{tabular}
\end{center}
\caption{The D-brane configuration}\label{Dconfig}
\end{table}

Note that there is no spatial direction along which all  the D-branes extend. Thus the low energy dynamics is that of an effective particle.
This feature is rather distinctive, as very often the low energy dynamics of microscopic descriptions of black hole in string theory is captured by an effective string \cite{Strominger:1996sh, Maldacena:1997de}. 

\subsection{Field content and gauge symmetry}
Excitation modes of the system correspond to to the modes of open strings stretched between these branes. Further the massless modes are the most relevant ones for low energy physics. Such modes/fields can be arranged in supersymmetric multiplets. We shall use the 4 dimensional supermultiplets, dimensionally reduced to $1$ dimension, i.e. only time.

There are two different types of open strings: strings starting and ending on the same brane and strings starting and ending on different branes.
\begin{enumerate}
\item
For each brane, first kind of strings give rise to the field content of $\mathcal{N}=4$ super  Yang Mills theory, which has a $\mathcal{N}=1$ vector multiplet and three $\mathcal{N}=1$ chiral multiplets. The bosonic content of these multiplets are summarized in Table \ref{adjfield}:
\begin{table}[htb]
\begin{center}
\begin{tabular}{|c||c|c|c|c|c|c|c|c|c|c}
\hline
& 1 & 2 & 3 & 4 & 5 & 6 & 7 & 8 & 9 \\
\hline
D2 & $X_1^{1}$ & $X_2^{1}$ & $X_3^{1}$ & Re $\Phi^1_1$ & Im $\Phi^1_1$ & Re $\Phi^1_2$ & Im $\Phi^1_2$ &  Re $\Phi^1_3$ &  Im $\Phi^1_3$ \\
\hline
D2 & $X_1^{2}$ & $X_2^{2}$ & $X_3^{2}$ & Re $\Phi^2_1$ & Im $\Phi^2_1$ & Re $\Phi^2_2$ &  Im $\Phi^2_2$ & Re $\Phi^2_3$ & Im $\Phi^2_3$ \\
\hline
D2 & $X_1^{3}$ & $X_2^{3}$ & $X_3^{3}$ & Re $\Phi^3_1$ & Im $\Phi^3_1$ &  Re $\Phi^3_2$ &  Im $\Phi^3_2$ & Re $\Phi^3_3$ & Im $\Phi^3_3$ \\
\hline
D6 & $X_1^{4}$ & $X_2^{4}$ & $X_3^{4}$ & Re $\Phi^4_1$ & Im $\Phi^4_1$ & Re $\Phi^4_2$ &  Im $\Phi^4_2$ & Re $\Phi^4_3$ & Im $\Phi^4_3$ \\
\hline
\end{tabular}
\end{center}
\caption{Bosonic fields coming from strings starting and ending on the same D-brane}\label{adjfield}
\end{table}
The superscripts denote the stack number. $(X_1^{k}, X_2^{k} , X_3^{k})$ describe the position of the $k^{th}$ brane in 3 dimensional non-compact space. $\Phi^k_{1,2,3} = \text{Re} \Phi^k_{1,2,3} + i \, \text{Im} \Phi^k_{1,2,3}$ respectively describe position of the $k^{th}$ brane along compact 4-5, 6-7 and 8-9 directions\footnote{These directions are compact, but as long as separation of diffent D-branes along  these directions are small, effects of compactness will not be relevant.}.

A stack of $N$ D-branes comes with a $U(N)$ gauge symmetry. Thus the three D2 branes are associated with three different $U(1)$ symmetries, whereas the D6 brane stack is associated with a $U(N)$ symmetry. Fields described above furnish adjoint representations of respective gauge groups. Since $U(1)$ acts trivially on adjoints, this means the fields on first three rows of Table \ref{adjfield} are gauge invariant or scalars, whereas those on the last row are $N \times N$ matrices transforming in adjoint representation of $U(N)$. 

\item
The second kind of strings for every pair $(kl)$ of branes  give rise to $\mathcal{N}=2$ hypermultiplet, or equivalently two $\mathcal{N}=1$ chiral multiplets $Z^{kl}, Z^{lk}$. We will abuse notation to denote the complex scalars of these multiplets by the same name. 
$Z^{kl}$ transforms as fundamental under $U(N_k)$ and anti-fundamental under $U(N_l)$. Here $N_k=1$ for $k=1,2,3$ and N for k=4. 
\end{enumerate}
Note, no field is charged under the overall $U(1)$. Hence the gauge symmetry of the system is $U(1) \times U(1) \times U(N)$, where  the two $U(1)$-s can be taken to be any two relative $U(1)$.

A D-brane system spontaneously breaks various translational symmetries of the ambient spacetime. These broken symmetries appear as Goldstones in the worldvolume theory of the D-branes. For the system  at hand these are
\begin{align}
\sum_{k=1}^3 \vec{X}^k + \frac{1}{N} \Tr{} \vec{X}^{4} \, , \, \Phi^1_1 + \frac{1}{N} \Tr{} \Phi^{4}_1  \, , \, \Phi^2_2 + \frac{1}{N} \Tr{} \Phi^{4}_2  \, , \, \Phi^3_3 + \frac{1}{N} \Tr{} \Phi^{4}_3  \, , \, \Phi^1_2 + \Phi^3_2 \, , \, \Phi^1_3 + \Phi^2_3 \, , \, \Phi^2_1 + \Phi^3_1 \, . \label{Goldstone}
\end{align}
The 7 supermultiplets containing these fields decouple from the dynamics. Fermions in these multiplets, which account for 28 off-shell degrees of freedom, correspond to the broken supersymmetries. This perfectly matches the fact that system preserves 4 real supercharges ($\mathcal{N}=1$ in 4 dimensional language) out of total 32 ($\mathcal{N}=8$ in 4 dimensional language).

Among the $\Phi$ fields, we can use translational symmetry to set $\Phi^1_1 = 0, \Phi^2_2 = 0, \Phi^3_3 = 0$. Then among various $\Phi$ fields, only the following combinations appear in the dynamics:\\
\begin{align}
\Phi^{(12)} &:= \Phi^1_3 - \Phi^2_3 \, , \, \Phi^{(23)}  = \Phi^2_1 - \Phi^3_1 \, , \, \Phi^{(31)} :=  \Phi^3_2 - \Phi^1_2 \, , \,  \Phi^{4}_1 \, , \, \Phi^{4}_2 \, , \, \Phi^{4}_3 \, . \label{phisurvive}
\end{align}

\subsection{The potential or loss function}
The detailed Lagrangian of the system is discussed in \cite{Chowdhury:2014yca, Chowdhury:2015gbk}. Here we directly jump to the potential. The potential (to be treated as the loss function) is a sum of 3 non-negative terms 
\begin{align}
V &= V_{gauge} +V_D + V_F \, .
\end{align}
$V$ is minimized, when each of $V_{gauge} ,V_D , V_F$ individually vanish\footnote{For spontaneously broken supersymmetry, minimum value of the potential occurs at some positive value of the potential. This is not the case here.}.
Among  these, $V_{gauge}$ can be set to 0, simply by putting the $X$ fields to zero, i.e. taking  3 non-compact coordinates of different stacks of D-branes to coincide. $X$ fields do not appear in $V_D$ and $V_F $, and hence can be safely forgotten. In fact, for non-zero values of $Z$ fields, which is the case for all minima, this is the only way to make $V_{gauge}$ vanish.

Setting $V_D$ to zero has the effect of fixing the scale of various fields. This effect can be incorporated by complexifying the $U(1) \times U(1) \times U(N)$ gauge invariance to $\mathbb{C}^* \times \mathbb{C}^* \times GL(N, \mathbb{C})$. Hence  it suffices to minimize $V_F$, subject  to the complexified gauge invariance. Thus in the following we shall discuss only about $V_F$ and relegate the details of $V_{gauge}$ and $V_D$ to Appendix \ref{VgaugeD}. 

The ``F-term potential'' $V_F$ is derived in terms of an underlying superpotential $W$, which is detailed in  Appendix \ref{appsup}.
For now, it is useful to first define some intermediate objects:
\begin{align}
\nn
F^{12} &=  Z^{12} Z^{21} + c^{12}   ~,~
F^{21} =  Z^{21} Z^{12} + c^{12} , \\
\nn
F^{23} &= Z^{23} Z^{32} + c^{23}  ~,~ 
F^{32} =  Z^{32} Z^{23} + c^{23}  , \\
\nn
F^{13} &=  Z^{13} Z^{31} + c^{13}   ~,~
F^{31} =  Z^{31} Z^{13} + c^{13}  , \\
\nn
F^{14} &= Z^{14}. Z^{41} + c^{14} N  ~,~
F^{41} =  Z^{41} Z^{14} + c^{14} \mathbb{I}_{N} + [\Phi_2^4, \Phi_3^4]  , \\
\nn
F^{24} &=  Z^{24}. Z^{42} + c^{24} N ~,~ 
F^{42} =  Z^{42} Z^{24} + c^{24} \mathbb{I}_{N} + [\Phi_3^4, \Phi_1^4]  , \\
\nn
F^{34} &=   Z^{34}. Z^{43} + c^{34} N  ~,~
F^{43} =  Z^{43} Z^{34} + c^{34} \mathbb{I}_{N} - [\Phi_2^4,\Phi_1^4] , \\
\nn
G^{21} &=  \Phi^{(12)}  Z^{21} + Z^{23} Z^{31} + Z^{24} . Z^{41}  ~,~ G^{12} =  \Phi^{(12)}  Z^{12} + Z^{13} Z^{32} + Z^{14} . Z^{42}  \, , \\
\nn
G^{31} &=  \Phi^{(31)} Z^{31}  + Z^{32} Z^{21} - Z^{34} . Z^{41}   ~,~ G^{13} =  \Phi^{(31)}  Z^{13}  + Z^{12} Z^{23} + Z^{14} . Z^{43} \, ,  \\
\nn
G^{32} &=  \Phi^{(23)} Z^{32}  + Z^{31} Z^{12} + Z^{34} . Z^{42}  ~,~ G^{23} = \Phi^{(23)} Z^{23} + Z^{21} Z^{13} + Z^{24} . Z^{43}  \, ,  \\
\nn
G^{41} &=   - \Phi_1^4 Z^{41} + Z^{42} Z^{21} + Z^{43} Z^{31}  ~,~ G^{14} =  - Z^{14} \Phi_1^4 + Z^{12} Z^{24} - Z^{13} Z^{34}  \, ,  \\
\nn
G^{42} &=  - \Phi_2^4 Z^{42} + Z^{43} Z^{32} + Z^{41} Z^{12}  ~,~ G^{24} = - Z^{24} \Phi_2^4 + Z^{21} Z^{14} + Z^{23} Z^{34}  \, ,  \\
G^{43} &=  - \Phi_3^4 Z^{43} - Z^{41} Z^{13} + Z^{42} Z^{23}   ~,~ G^{34} =  - Z^{34} \Phi_3^4 + Z^{31} Z^{14} + Z^{32} Z^{24}  \, . \label{Fterm}
\end{align}
Here $c^{12}, c^{13}, c^{23}, c^{14} , c^{24} , c^{34}$ are complex non-zero parameters. Note $F^{41}, F^{42}, F^{43}$ are matrices, $G^{41}, G^{42}, G^{43}$ are column vectors, $G^{14}, G^{24}, G^{34}$ are row vectors and rest of the combinations are numbers. The F-term potential $V_F$ is the  sum of modulus square of 
all terms in \refb{Fterm}, i.e. 
\begin{align}
\nn
V_F &=  \sum_{i \neq j, i,j =1}^3 |F^{ij}|^2 + \sum_{i=1}^3 |F^{i4}|^2 +  \sum_{i=1}^3 \Tr{} (F^{4i})^\dagger F^{4i} \\
&+ \sum_{i \neq j, i,j =1}^3 |G^{ij}|^2 + \sum_{i=1}^3 \left( (G^{4i})^\dagger G^{4i} + G^{i4} (G^{i4})^\dagger  \right) \label{V_F}.
\end{align}
We will use the words minima and solutions interchangeably.
It suffices to think of $V_F$ as the loss function. Minima of $V_F$ are solutions to the equations
\begin{align}
F^{kl} &= 0 \, , \, G^{kl} = 0 \, , \quad k \neq l, \, k,l =1,2,3,4 \, . \label{Feqn}
\end{align}
Equations \refb{Feqn} are invariant under the complexified gauge transformations
\begin{align}
\nonumber
&Z^{12} \rightarrow \lambda_{12} Z^{12} \, , \, Z^{21} \rightarrow \lambda_{12}^{-1} Z^{21} \, , \, 
Z^{23} \rightarrow \lambda_{23} Z^{23} \, , \, Z^{32} \rightarrow \lambda_{23}^{-1} Z^{32} \, , \\
\nonumber
&Z^{13} \rightarrow \lambda_{12} \lambda_{23} Z^{13} \, , \, Z^{31} \rightarrow \lambda_{12}^{-1} \lambda_{23}^{-1} Z^{31} \, , \\
\nonumber
&Z^{14} \rightarrow \lambda_{12} \lambda_{23} Z^{14} M^{-1} \, ,  Z^{24} \rightarrow \lambda_{23} Z^{24} M^{-1} \, ,  Z^{34} \rightarrow  Z^{34} M^{-1} \, , \\
\nonumber
&Z^{41} \rightarrow \lambda_{12}^{-1} \lambda_{23}^{-1} M Z^{41} \, , \, Z^{42} \rightarrow \lambda_{23}^{-1} M Z^{42} \, , \, Z^{43} \rightarrow M Z^{43} \, , \\
&\Phi_i^{4} \rightarrow M \Phi^{4}_i M^{-1} \, . \label{complexgauge}
\end{align}
where $\lambda_{12} \in \mathbb{C}^\ast, \lambda_{23} \in \mathbb{C}^\ast,  M \in GL(N, \mathbb{C})$. The $U(1) \times  U(1) \times U(N)$ subgroup of this is the gauge group.  

Solutions to \refb{Feqn}, that are related by $ \mathbb{C}^\ast \times  \mathbb{C}^\ast \times  GL(N, \mathbb{C})$ gauge transformations acting as in \refb{complexgauge}, are to be counted as same solution. In other words, a physical solution really stands for a gauge orbit.
In this work, we shall get rid of $ \mathbb{C}^\ast \times  \mathbb{C}^\ast \times  GL(N, \mathbb{C})$ redundancy, by fixing gauge and will minimize the  loss function in remaining variables. Various choices of gauge are discussed in Appendix \ref{appgauuge}.

If the gauge redundancy is not fixed, then each discreet  minima turns into a continuous space of minima, corresponding to the gauge orbit. This can be thought of as a trivial realization of 
is similar to the phenomena of mode connectivity found in some NN architectures, i.e. existence of paths of almost constant loss connecting two minima of the loss function  \cite{garipov2018loss, raghavan2020sparsifying}.


The minima of $V_F$ can also be thought of as minima of a slightly simpler potential \refb{Vfsimple}. We relegate this to Appendix \ref{appsimple}, since this simpler potential  \refb{Vfsimple} is not the physical potential. But if one's sole concern is to find a potential landscape with exponentially many minima, then this simpler potential is as good as  the physical one. For the special  case of $N=1$, the system can be further simplified to one with only 3 complex variables. This is discussed in Appendix \ref{appN1}.

\section{Searching for multiple minima with Stochastic  Gradient Descent} \label{s4}
Different minima of the loss function represent different ways of learning the same data, with minima having lower loss value representing better learning. When all the minima are global, as in linear DNN-s \cite{kawaguchi2016deep}, all minima ought to be equally good by this standard. However, since the weights and biases of NN-s corresponding to different global minima are different, something must be different about learnings corresponding to different minima. 
Whether this difference is merely technical or it translates to some qualitative differences is not obvious at this stage. It is conceivable that at least some different minima might actually differ qualitatively. E.g. wider minima might offer better generalisability \cite{NIPS1994_01882513, chaudhari2017entropysgd, keskar2017largebatch}.

To study this important issue, first of all we need to be able to find all (or at least a large fraction of all) the minima. 
The ability of an algorithm to find various minima thus becomes relevant. 
The black holes discussed in last section provide an excellent settings to test this.
In this section, we discuss how  SGD performs in this regard, i.e. whether or not and in how many runs it can find all the minima. 
\subsection{Estimated number of runs required to find all the minima}
Before we set out, it is useful to estimate the typical number of runs in which one might expect to find all the minima. More precisely, 
\textit{ if the loss function has $K$ minima and an algorithm randomly  hits one minima in each run with equal likelihood, then what is the probability $p(K,n)$ of finding all the solutions after $n$ runs?}

Such a $p(K,n)$ must satisfy
\begin{align}
\nn
p(K,n) &= 0 \quad \text{for} \quad n=1, \dots , K-1 \, , \\
\lim_{n \rightarrow \infty} p(K,n) &= 1 \, . \label{pexpect}
\end{align}
Running the algorithm $n$ times will produce a sequence (of minima) of length $n$. There are $K^n$ such sequences. The question at hand can be phrased as combinatorial questions about such sequences and the answer will likely involve objects like binomial coefficients. 
We make the following ansatz for $p(K,n)$
\begin{align}
p_{ansatz} (K,n) &:= 1 - \frac{1}{K^n} \sum_{a=1}^{K-1} (-1)^{a-1} ~ {{K} \choose {a}}  \,  (K-a)^n \, , \label{pKn}
\end{align}
which satisfies the condition \refb{pexpect} $\forall \, n, K \geq 1, n, K \in \mathbb{Z}$. Here ${{K} \choose {a}} = \frac{K!}{ a! (K-a)!}$ are the binomial coefficients\footnote{sometimes denoted as $~^KC_a$.}. Satisfaction of the second condition of \refb{pexpect} is obvious, whereas satisfaction of the first condition can be verified. For the simple case of K=3, a derivation of this formula is given in Appendix \ref{appp3n}.

\begin{figure}[H]
\begin{center}
\includegraphics[scale=.8]{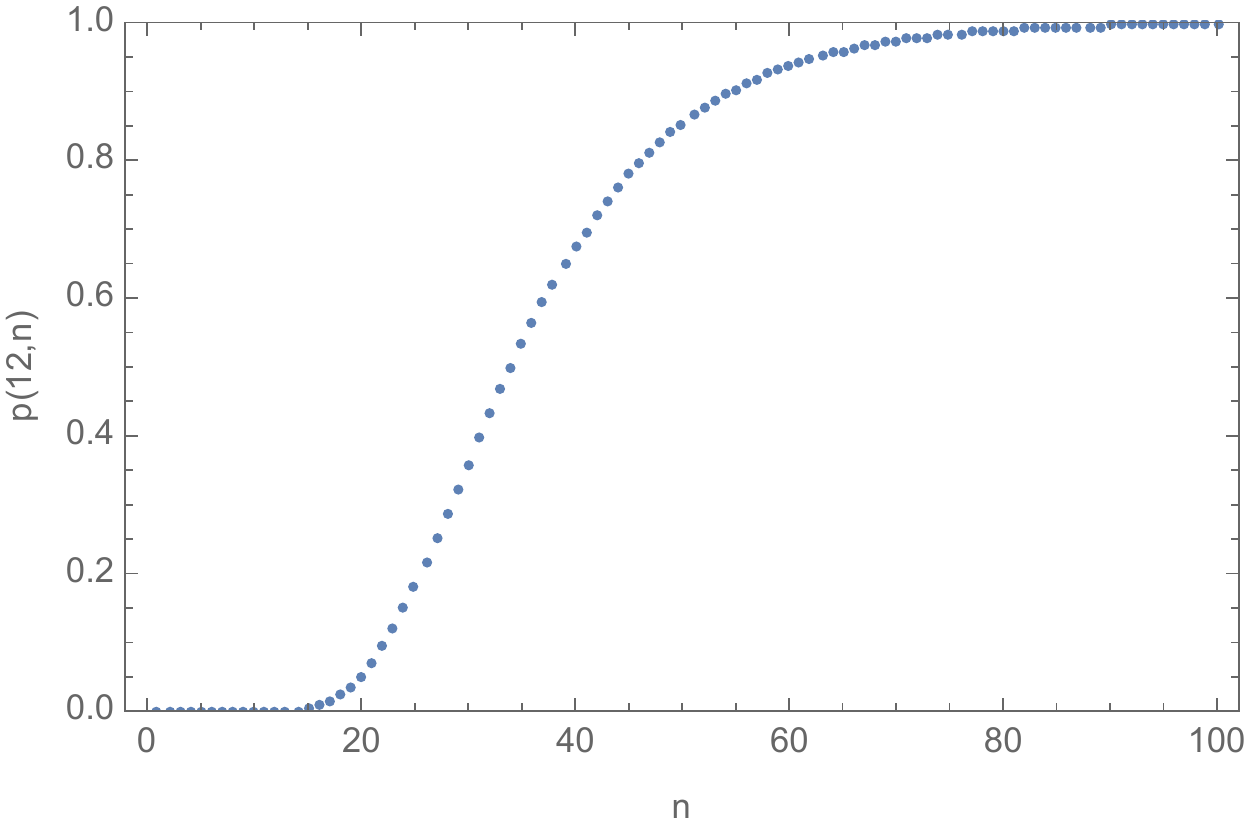}
\caption{Probability $p(12,n)$ of finding all 12 minima in $n$ runs}\label{run12}
\end{center}
\end{figure}

For $K=12$, which is the number of minima of \refb{V_F} for $N=1$, the function $p(K,n)$ is plotted in Fig. \ref{run12}.
Note, the estimate \refb{pexpect} ignores the possibility that a run may fail to find any minima whatsoever. As we discuss later, we actually find this to be a frequent occurrence.

A measure of efficiency of an algorithm in finding all the minima might be introduced as follows. 
 Let $n_\epsilon$ be the number of runs needed to reach probability $1 - \epsilon$ for some small $\epsilon$, i.e.  $p(K, n_\epsilon) = 1-\epsilon$.
Let the fraction of minima found by an algorithm after $n_\epsilon$ runs be $f (n_\epsilon)$. Then we might define 
\begin{align}
\mathbb{E}_\epsilon := \frac{f (n_\epsilon)}{1- \epsilon} \, . \label{effdef}
\end{align}
As the number of runs approach infinity, or equivalently $\epsilon \rightarrow 0$,  the efficiency $\mathbb{E}_\epsilon$ approaches its asymptotic value $\mathbb{E}_0$. If the algorithm is unable to find all the minima, then $\mathbb{E}_\epsilon<1$ for small enough $\epsilon$.
If the algorithm is able to find all the minima after $n_c$ runs, then $\mathbb{E}_{\epsilon(n_c)} >1$. If one keeps running the algorithm, $\mathbb{E}_\epsilon$ approaches its asymptotic value $\mathbb{E}_0 = 1$.

\subsection{Black holes: A case study}

\subsubsection{$N=1$}

The case $N=1$ is the simplest one and admits 12 global minima. The minima can be found using Mathematica \cite{Mathematica} at one go, and rather quickly. The goal here is however to check whether the SGD can at all find all 12 minima or not. This question is of some interest since to the best of our knowledge, it is not clear as of now whether or  not SGD can find all the minima of a loss function. The tensorflow library \cite{tensorflow2015-whitepaper} is particularly helpful for this task. Note that although we are using the machinery of machine learning, we are not quite doing machine learning. In particular, there is no neural network involved.

Coming to the loss function, firstly we note that if the parameters $c_{ij}$-s are all taken to be real, then the equations \refb{Feqn} and the loss function \refb{V_F} remain unchanged under simultaneous complex conjugation and/or sign change of all the fields. This implies that given any minima of the loss function \refb{V_F} a field configuration obtained by complex conjugation and/or sign change of a minima, should also be a minima. Thus the minima occur in quadruplets. In the special case, where all the fields in a minima are either real of imaginary, these operations give doublets. To exploit these symmetries, we consider real $c_{ij}$-s. To be specific, we work with the following set  of $c_{ij}$ parameters:
\begin{align}
c_{12} &= 2/3, \, c_{31} = 3/5, \, c_{14} = 5/7, \, c_{23} = 7/11, \, c_{24} = 11/13, \, c_{34} = 13/17 \, . \label{c}
\end{align}
There is a $\mathbb{C}^\ast \times \mathbb{C}^\ast \times \mathbb{C}^\ast$ complexified gauge redundancy, which affects various Z-fields, but not the $\Phi$-fields. This gauge redundancy can be fixed for example by fixing $Z^{12}, Z^{13}, Z^{14}$ to arbitrary complex numbers. The specifics of gauge choice affect the minima, but not the gauge invariant fields or combinations of fields. Thus the gauge invariants offer a useful way of labelling the  minima.  In Table \ref{solnN=1}, we  mention the $\Phi$-fields  for all 12 minima obtained using Mathematica \cite{Mathematica}. By virtue of being  gauge invariant, any choice of gauge will result in $\Phi$-fields for minima to be those in Table \ref{solnN=1}.
\begin{table}[htb]
\begin{center}
\scalebox{0.6}{
\begin{tabular}{|c||c|c|c|c|c|c|c}
\hline
~ & $\Phi^{(12)}$ & $\Phi^{(23)}$ & $\Phi^{(31)}$ & $\Phi^{4}_1$ & $\Phi^{4}_2$ & $\Phi^{4}_3$ \\
\hline
1  & 0.114667 & 3.04279 & -0.426835 & 3.07604 & 0.00561356 & 0.441079 \\
 \hline
2 & -0.114667 & -3.04279 & 0.426835 & -3.07604 & -0.00561356 & -0.441079 \\
\hline
3  & 2.50319 & 0.148546 & -0.475075 & 0.523547 & -0.00663202 & 2.54768  \\
 \hline
4 & -2.50319 & -0.148546 & 0.475075 & -0.523547 & 0.00663202 & -2.54768 \\
 \hline
5 & 0.757429 & 0.703097 & -0.769789 & 1.05802 & -0.181488 & 1.06349 \\
 \hline
6 & - 0.757429 & -0.703097 & 0.769789 & -1.05802 & 0.181488 & -1.06349\\
 \hline
7 &  - 0.113796  i &  - 0.126033  i &  + 2.48169  i &  0.653677  i &  2.40634  i &  0.616251  i \\
 \hline
8 &  0.113796  i &  0.126033  i &  - 2.48169  i &  -  0.653677  i &  - 2.40634  i &  - 0.616251  i  \\
 \hline
9 & 0.133727 + 1.5199  i & 0.139324 + 1.60413  i & -0.966204 + 0.664837  i & 0.829512 - 0.761908  i & -0.0896705 - 1.37815  i & 0.77546 -  0.721537  i \\
 \hline
10 & 0.133727 - 1.5199  i & 0.139324 - 1.60413  i & -0.966204 -  0.664837  i & 0.829512 + 0.761908  i & -0.0896705 + 1.37815  i & 0.77546 +  0.721537  i \\
  \hline
11  & -0.133727 + 1.5199  i & -0.139324 + 1.60413  i & 0.966204 +  0.664837  i & -0.829512 - 0.761908  i & 0.0896705 -  1.37815  i & -0.77546 - 0.721537  i \\
 \hline
12 & -0.133727 - 1.5199  i & -0.139324 - 1.60413  i & 0.966204 -  0.664837  i & -0.829512 + 0.761908  i & 0.0896705 +  1.37815  i & -0.77546 + 0.721537  i \\
 \hline     
\end{tabular}}
\end{center}
\caption{Minima for N=1}\label{solnN=1}
\end{table}

\begin{figure}[H]
\begin{center}
\includegraphics[scale=1]{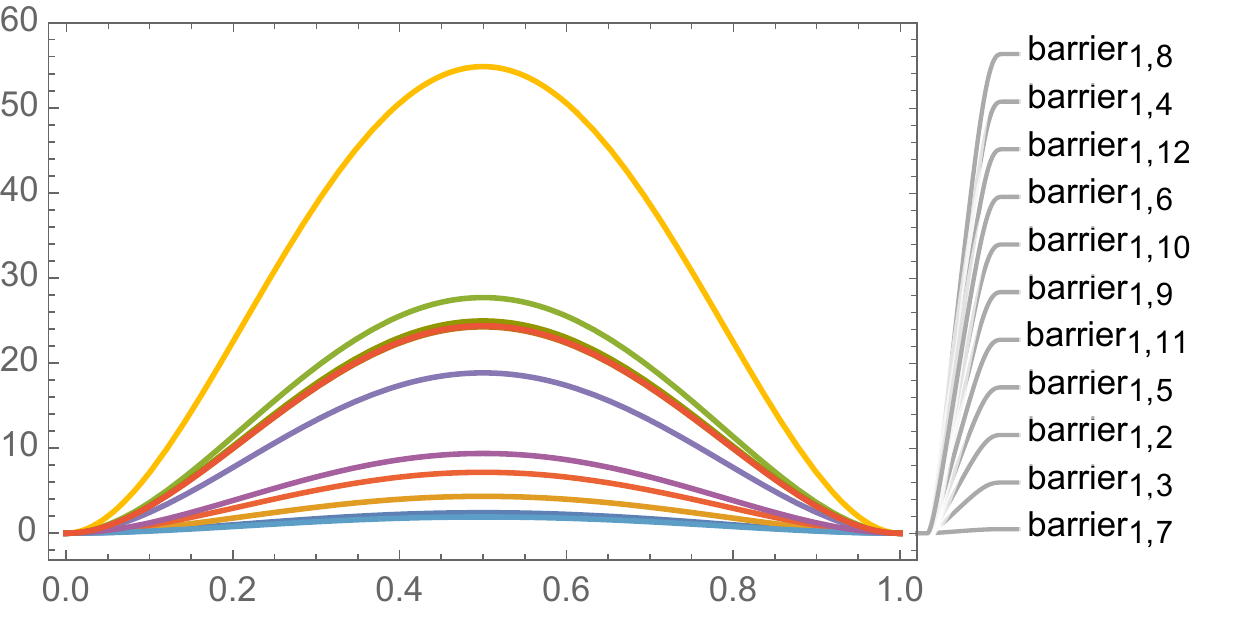}
\caption{Potential barriers along a straight line connecting minimum 1 to other 11 minima for  $N=1$. The x-axis parameterises a line segment connecting solution 1 to solution $j$ for $j=2, \dots , 12$, such that $x=0$ corresponds to solution 1 and  $x=1$ corresponds to solution $j$, ``barrier$_{1,j}$'' denotes the potential along such a line segment.}\label{hill}
\end{center}
\end{figure}

One can further proceed to obtain some understanding of the loss function landscape. We only undertake some preliminary exploration  in this direction. Given that we have the analytic form of the loss function \refb{V_F} as well as locations of the minima, one might wonder about mode connectivity, i.e. existence of paths of relatively low loss connecting these minima. With this  question in mind we study the loss function along line segments connecting different minima. The resulting cross section of the loss function vanishes at the ends of such segments, and rises to some maximal value in the middle\footnote{Empirically we find loss function along these segments to be highly symmetric, although the reason behind this symmetry is not apparent to us.}. This maximal value varies depending on the segments. E.g. in Figure \ref{hill}, we plot the loss function along segments connecting minimum 1 to other 11 minima. For some of the segments, the maximal value of the loss function is rather low. This feature resembles mode connectivity. The maximal value presumably depends on the parameters $c_{ij}$-s. It is unclear whether mode connectivity persists even if  $c_{ij}$-s are taken to be larger than those in \refb{c}.
Other questions of interest include distribution of critical points, especially the saddle points. We do not attempt this here. 

Now we comment  on number of runs needed to find the minima. If we use the information that the minima appear in doublets/quadruplets, then it suffices to find one minima in every doublet/quadruplet. We could find at least one minima in each doublet/quadruplet in 35 runs. 
Thus the efficiency $\mathbb{E}_0$, as defined in  \refb{effdef} is unity.
The number 35 is significantly smaller  than the number of estimated runs, a random algorithm would take to find all the minima \refb{pKn}. Figure \ref{run12} suggests it would take about 80 runs to find all 12 minima. 

However the situation worsens, if we forget about the symmetries and try to find all 12 minima independently. 
Counting  configurations with loss value $< 10^{-7}$ as valid  minima, we have only 11 out of 12 minima in 200 runs\footnote{Actually we did get the 12-th  one (first minima in Table  \ref{runN=1}), but with a loss value around $10^{-3}$, so it could not be counted.}. These accounted for 109 runs. Among the remaining runs, some reached comparatively larger but still small loss values and the configurations were close to various minima.
To be specific, number of configurations reached with loss of the order of $10^{-7}, 10^{-6}, 10^{-5}, 10^{-4}, 10^{-3}, 10^{-2}$ respectively are $2,1,5,3,5,3$. 
Rest had even higher loss values and some showed runaway behaviour as well.

Curiously, different minima appear with different frequencies. Table \ref{runN=1} summaries which minima was hit how many times. 
\begin{table}[htb]
\begin{center}
\scalebox{1.0}{
\begin{tabular}{|c|c|}
\hline
minima number & number of hits \\
\hline
1   & 0   \\
 \hline
2   & 1 \\
\hline
3   &  8  \\
 \hline
4   &  8   \\
 \hline
5   & 24  \\
 \hline
6   & 25  \\
 \hline
7   & 3  \\
 \hline
8   &  3  \\
 \hline
9  & 6  \\
 \hline
10   & 12  \\
  \hline
11   & 10  \\
 \hline
12  & 9 \\
 \hline     
\end{tabular}}
\end{center}
\caption{Number of times various minima for N=1 was found (out of 200 runs), i.e. loss function dipped below $10^{-7}$}\label{runN=1}
\end{table}
We note that minima in same doublet/quadruplet often appear with same/similar frequency, with minima 5,6 being the most frequent and minima 1,2 being the list. This raises the question -  what is special about  the minima that are found more frequently? We note that the  least frequent minima hosts the largest as well as the smallest Hessian eigenvalues among all minima\footnote{Hessian eigenvalues are the same for minima in same doublet/quadruplet.}, and therefore has the widest range of eigenvalues. The  Hessian eigenvalues of various doublet/quadruplet are given in Table \ref{HessianN=1}. 
%
%
\begin{table}[htb]
\begin{center}
\scalebox{0.5}{
\begin{tabular}{|c||c|c|}
\hline
minima number  & eigenvalues of the Hessian (up to 4 decimal points) \\
\hline
1,2  & 290.331, \, 290.331, \,49.0113, \,49.0113, \,32.591, \,32.591, \,16.29, \,16.29, \,2.98038, \,2.98038, \, 2.21897, \,2.2189, \,
    1.5384, \,1.5384, \,0.0551, \,0.0551, \,0.0004, \,0.0004   \\
\hline
3,4  &  195.306, \, 195.306, \, 38.7135, \, 38.7135, \, 20.3297, \, 20.3297, \, 14.1516, \, 
14.1516, \, 3.20372, \, 3.20372, \, 2.1753, \, 2.1753, \, 1.3116, \, 1.3116, \, 
0.0678, \, 0.0678, \, 0.0007, \, 0.0007 \\
 \hline
5,6  & 31.6017, \,31.6017, \,14.2884, \,14.2884, \,8.6984, \,8.6984, \,4.4307, \,4.4307, \,2.8221, \, 2.8221, \,2.0027, \,2.0027, \,1.1111, \,1.1111, \,0.0494, \,0.0494, \,0.0184, \,0.0184 \\
 \hline
7,8  & 99.8704, \,99.8704, \,31.1475, \,31.1475, \,13.7417, \,13.7417, \,9.9342, \,9.9342, \,3.1265, \, 3.1265, \,2.4533, \,2.4533, \,1.3331, \,1.3331, \,0.1656, \,0.1656, \,0.0110, \,0.0110 \\
 \hline
9,10,11,12  &  25.8232, \,25.8232, \,9.8151, \,9.8151, \,5.1493, \,5.1493, \,3.4729, \,3.4729, \,3.0792, \,    3.0792, \,2.412, \,2.412, \,1.8277, \,1.8277, \,0.5702, \,0.5702, \,0.2699, \,0.2699  \\
 \hline
\end{tabular}}
\end{center}
\caption{Eigenvalues for Hessians of different doublets/quadruplets of minima for N=1}\label{HessianN=1}
\end{table}

\begin{figure}[H]
\begin{center}
\includegraphics[scale=.5]{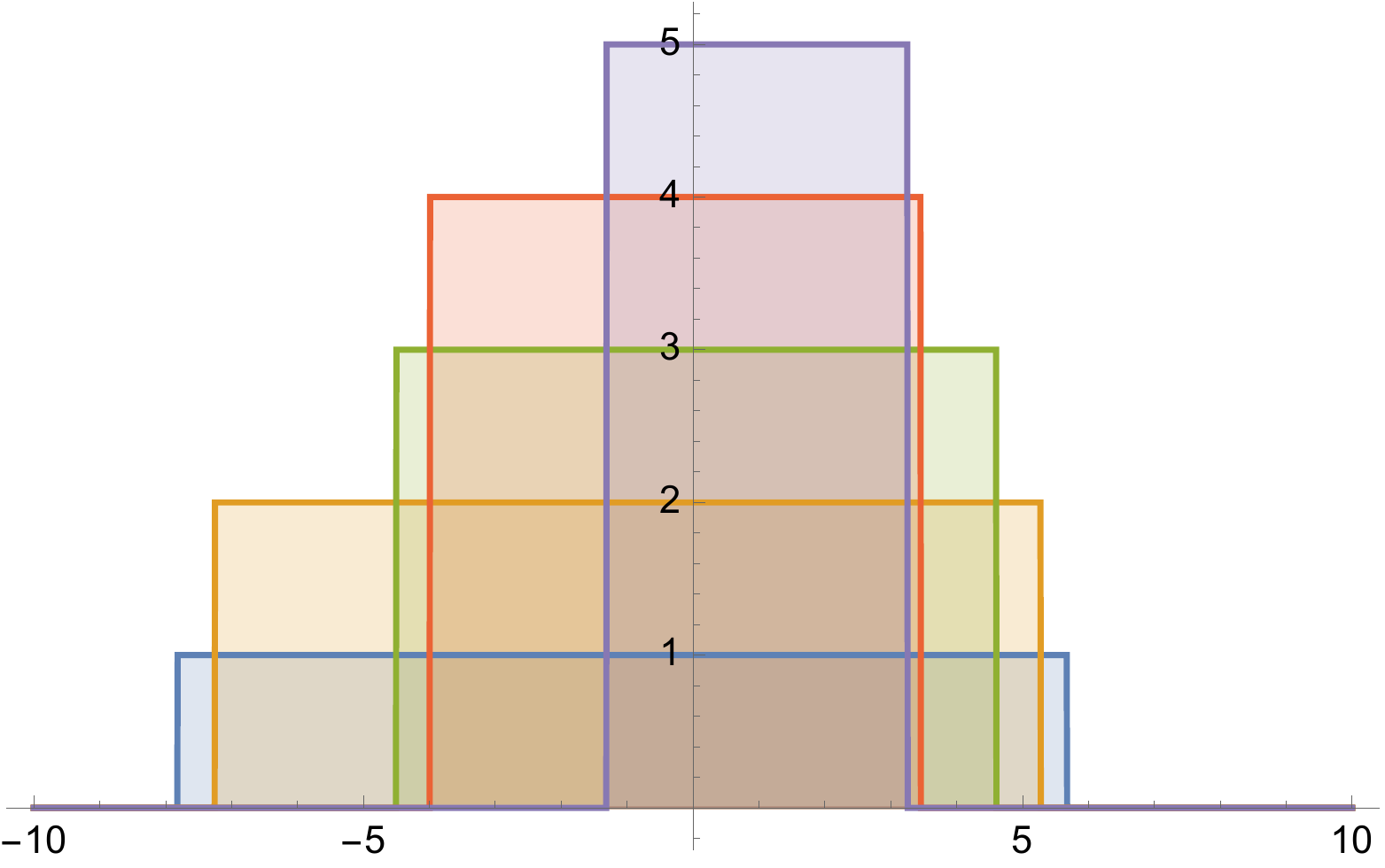}
\caption{A plot of logarithm of the eigenvalues ranges}\label{hess}
\end{center}
\end{figure}

A visual representation of the same is given in Figure \ref{hess}. Five blocks represent five different eigenvalue sets. We choose to plot the logarithms, since the eigenvalues differ by several orders of magnitudes. The widest block at the  bottom of Figure \ref{hess} corresponds to minima 1,2. The narrowest block, corresponding to minima 9,10,11,12, however are not the most frequent ones.

%

\subsubsection{$N>1$}
The loss function \refb{V_F} becomes significantly complicated for $N>1$ due to non-Abelian nature of the variables. But  the essential task remains the same. The loss function \refb{V_F} vanishes at the global minima and we decide on some small cut-off for loss function, which  if reached, we take the configuration to be a minima. We take this cut-off to be $10^{-6}$. 

Again, using the fact that minima appear in doublets/quadruplets, it suffices to find one minimum for each doublet/quadruplet. With this fact utilised, for the $N=2$ case we could find all 56 of 56 minima, whereas for the $N=3$ case we could find 176 out of 208 minima. The efficiency $\mathbb{E}_0$ is respectively $1$ and $11/13 \sim .846$ for $N=2,3$.
Number of runs in both cases were were hopelessly high and hence we did not keep track of the same. 

For $N=4$, we were unable to reach  small enough values of the loss function in reasonable time using SGD. We believe this obstacle can be overcome with more effort and perhaps more expertise. But since is is not indispensable for the main argument of this work, we do not attempt it here.

As in the case of $N=1$, we use SGD to look for the minima. We have made extensive use of the Tensorflow library \cite{tensorflow2015-whitepaper}. As previously stressed, we are not machine learning anything and there is no neural network in sight. We are merely using the machinery of machine learning to find the minima of \refb{V_F}. 

Due to large number of variables involved, we do not list the field configuration corresponding to the minima for $N>1$. The interested reader is referred to \cite{Chowdhury:2015gbk} for the case $N=2$.

In the following, we mention some relevant details of our quest for the loss minima. Many of these details are also present in the $N=1$ case, but assume more significance for $N>1$, due to increased complexity.

\begin{itemize}
%

\item
Initial values: \,
Since the run might hit different  minima depending on the starting point, we chose the initial values of all the variables randomly in the range (-5,5) from a normal or uniform distribution. For initial values beyond this range, the runs usually manifested a runaway behaviour, presumably due to the quartically diverging nature of the loss function. We also considered purely real/imaginary initial values, which yielded only limited success. 

We also explored some special initial values, where derivatives of the loss function along some directions vanish and also initial values at which some of the 
$F$-s and $G$-s  \refb{V_F}, \refb{Feqn} vanish. But  these did not lead to any new minima.


\item
Two Cycles Optimization: \,
As many runs seem  to get stuck at plateaus with
 loss value of order 1, we switched to a two cycle optimization where we would only continue if the loss value had come below a set threshold after running for a set number of steps. If it had, we would use those values of parameters to run a second cycle, using that same optimizer, or a different one. 


\item
Hyperparameter Tuning: \,  
The most important hypermeter is the learning rate, which we took to be smaller than $10^{-2}$, as the probability of runaway behaviour increased above this value. The number of steps needed to reach a minimum was of the order of $10^5$ (sometimes more). Smaller learning rate (e.g. $< 10^{-3}$) required more steps (of the order of $10^7$ or more) and more time (order of hours) to find a minimum. However once a small enough loss value has been reached, smaller step-size is desirable. So sometimes we tweaked the learning rate in the middle of the run accordingly. 
Another important hyperparameter was momentum, whose default value of 0.9. Even with these precautions, roughly one in every ten runs converged to a minima for the case of $N=3$.

%
%
%
%

\item
Excluding already found minima: \, 
As in $N=1$, some minima are found more frequently than others for $N>1$ as well. To avoid this repetition, one may  modify the loss function by adding a hump around an already found minima $x_0$. This modified loss function does not have $x_0$ as a global minima, but other minima of  the original loss function continue to be the minima of the modified loss function. A smooth choice of hump function will affect the locations as well as values of other minima as well, but these effects can be made arbitrarily small by appropriate choice of the hump. A more serious problem occurs due to new local minima and/or saddles introduced by the hump function. Not surprisingly, we did not find such modification of loss landscape to be helpful.
A simple realization of this idea is presented in Appendix \ref{appexclude}. 

\item
Gauge choice: \,
For N=2 and higher, finding all the minima in a single gauge proved difficult, hence we ran the search for multiple gauge choices. For N=2, two gauge choices sufficed to produce all the minima, but more were required for N=3. In Appendix \ref{appgauuge}, we discuss a wide class of gauge choices for arbitrary $N>1$.

In order to distinguish same minima found in different gauges, gauge invariant markers were calculated and used to count the number of linearly independent minima found(given the choice of special $c_{ij}$-s, minima can have two or four fold symmetry). The simplest gauge invariants are $\Phi^{(12)}, \Phi^{(23)}, \Phi^{(31)}, \Tr{} \Phi^4_1, \Tr{} \Phi^4_2, \Tr{} \Phi^4_3$.

\item
Scaling: \, 
Under the scale transformations $Z \to \lambda Z, \, \Phi \to \lambda \Phi, \, c_{ij} \to \lambda^2 c_{ij}$, the loss function \refb{V_F} changes by an over all scaling factor of $\lambda^4$. It follows that the minima of this rescaled loss function are related to those of the original one by simple scaling.
This fact can be used to scale up/down the $c_{ij}$-s, find minima of the resultant loss function and then transform back the minima by reverse scaling. This essentially has the effect of zooming in/out the loss landscape. This technique led to some new minima for $N=3$.

\end{itemize}

\section{Discussion} \label{s5}
In this work we have focused on the presence of exponentially many low lying local minima of the loss landscape, which seems to be a generic feature found in diverse NN architectures. We have pointed out that the physical potential in microscopic description of black holes in string theory naturally give rise to similar landscapes.
This is quite striking since these potentials can have as few as a couple of dozens parameters in simple cases, compared to millions of parameters in realistic DNN models. Furthermore, in this case, the exact number of minima is known a priori, owing to the stringent mathematical structure of string theory. Physically, the presence of exponentially many minima has its origin in black hole entropy.
Apart from establishing a curious connection between quantum gravity and machine learning, our work provides a large class of computationally cheap testing grounds for studying questions related to the loss landscape, some of which we have explored in this work.


The possibilities offered by the connection made in this work far exceed explorations carried out here. 
Firstly, our computational resources limited our exploration to landscapes for small charges, i.e. landscapes with relatively small number of minima. It is imperative to carry out similar investigations for larger charges, where the exponential degeneracy of minima becomes apparent. We have mostly relied on SGD, as means of finding the minima, as preliminary experience indicated SGD to be the most efficient. However a detailed study of the performances of other algorithms is desirable. 

Apart from finding the minima, the landscapes discussed  in this work, offer perfect settings for testing the role of saddle points in search convergence. We have noted that with increasing $N$, the search gets slower. It would be interesting to enumerate the number of saddles, which should not be too difficult as the potential has an analytic form. Although we have not attempted this in the present work. 

More realistic parallels of loss landscapes of non-linear DNN-s have non-degenerate minima. As mentioned earlier, such landscapes can easily be obtained from the ones discussed in this work, simply by adding to the potential a slowly varying function bound from below.  It would be interesting to explore  whether and how such changes affect the search for minima.

A more futuristic, yet perhaps most intriguing possibility would be to machine learn the minima of the loss function themselves. At present times, when one is usually content with finding any one set of hyperparameters with low enough loss value, this may sound too far-fetched. But as the technical advancements take place at a  galloping pace, it may soon be commonplace to find several local minima of the loss function of a DNN. Questions pertaining to multitude of minima would then be of practical interest. In the meantime, we might obtain theoretical insights into similar questions by studying the potential landscapes pointed out in this paper, which are much simpler to study due to small number of parameters involved.

\paragraph{Acknowledgement:} We are indebted to Ashoke Sen, 
for pointing out that black hole microstate counting can be framed as a machine learning problem as well as for his comments. 
It is a pleasure to thank Shailesh Lal and Challenger  Mishra for illuminating discussions. The work of T.M. is supported by the grant SB/SJF/2019-20/08. The work of S.M. was supported by Dublin Institute for Advanced Studies.
\newpage
\appendix

\section{The index} \label{appindex}
The black hole under consideration preserves 1/8 of the full supersymmetry. This can be seen as follows: the  each stack of D-branes naively preserves half of the available supersymmetry. Applied to first three stacks, this implies preservation of 1/8 of total supersymmetry. It can be checked that the fourth stack does not break any remaining supersymmetry. 

Broken  supersymmetries lead to 32-4=28 Goldstinos in the theory describing the intersecting D-brane system. Since the D-brane system  nevertheless preserves some supersymmetry, there exists 28 Goldstones as well, given in \refb{Goldstone}.

The  naive Witten  index vanishes due to the Goldstinos. This is because any Goldstino will be associated with two fermionic creation operators, and given any state, leads to four degenerate states, two of which are bosonic and two fermionic.

Thus in the trace, one must insert extra factors to soak up the Goldstinos. This task is accomplished in this case by the 14-th helicity  supertrace \cite{Bachas:1996bp, Gregori:1997hi}
\begin{align}
B_{14} &= - \frac{1}{14 !}  \Tr{} (2J_3)^{14} (-1)^{2J_3} \, .
\end{align}
Computing $B_{14}$ is same as the computing Witten index in a reduced system, obtained by removing the Goldstone multiplets. 

In a different duality frame, these indices have been computed \cite{Shih:2005qf, Sen:2009gy}.
For configurations considered in this work, i.e. with charge vector $(1,1,1,N)$, 
 the generating function for $B_{14}(N)$ is given by 
\begin{align}
\sum_{N=0}^\infty B_{14}(N) q^N &= \frac{\sum_{n \in \mathbb{Z}} q^{n(n+1)}}{ \prod_{n=1}^\infty (1-q^n)^6} \, .
\end{align}
The right hand side can be easily expanded in Mathematica and the expansion coefficients can be extracted.
First few terms of this expansion are computed to be \cite{Mathematica}
\begin{align}
2+12 q+56 q^2+208 q^3+684 q^4+2032 q^5+5616 q^6+14592 q^7+36088 q^8+85500 q^9+195312 q^{10}+ \dots \, .
\end{align}

\section{Details of $V_{gauge}$ and $V_D $} \label{VgaugeD}
We  write $V_{gauge}$ and $V_D$ for the most general case where the $k^{th}$ stack, $k=1, 2 , 3 , 4$, has $N_k$ of D-branes. This puts all four stacks in same footing and enables usage of similar notations for all stacks, hence simplifying the expressions of $V_{gauge}$ and $V_D$. The case at hand can easily be obtained by putting $N-1=N_2=N_3=1, N_4=N$. 

The gauge potential is given by
\begin{align}
\nonumber
V_{gauge} &= \sum_{k,l=1; k\neq l}^4 \sum_{a=1}^3 \Tr{} \left[ \left( H^{kl}_a \right)^\dagger \left( H^{kl}_a \right) \right] + \sum_{k=1}^4 \sum_{a=1}^3 \sum_{A=1}^3 \Tr{} \left[ \left( Y^{k}_{aA} \right)^\dagger Y^{k}_{aA} \right] \\
\nonumber
&+ \frac{1}{4} \sum_{k=1}^4 \sum_{a,b=1}^3 \Tr{} \left[ \left( X^{k}_{ab} \right)^\dagger X^{k}_{ab}  \right] \, , 
\end{align}
where
\begin{align}
X^{k}_{ab} &= [X^{k}_a, X^{k}_b]\, , 
Y^{k}_{aA} := [X^{k}_a,\Phi^{k}_A] \, , H^{kl}_a := X_a^{k} Z^{kl} - Z^{kl} X_a^{l} 
\end{align}
F-term equations entail that all $Z$-s and $\Phi$-s are non-vanishing. Demanding a vanishing $V_{gauge}$ then implies
\begin{align}
X^{k}_a&= 0 \quad k=1,2,3,4, \, a=1,2,3 \, .
\end{align}
The D-term potential has the general form
\begin{align}
V_D &= \frac{1}{2} \sum_{k=1}^4 \Tr{} \left[ \left( \sum_{l \neq k} Z^{kl} Z^{kl \dagger} -  \sum_{l \neq k} Z^{lk \dagger} Z^{lk} + \sum_{A=1}^3 \left[ \Phi^{k}_A, \Phi^{k\dagger}_A \right] - c^{k} \mathbb{I}_{N_k} \right)^2 \right] \, .
\end{align}
$c^{(k)}$-s are arbitrary real numbers, called Fayet-Illipoulos (FI) parameters,  satisfying $\sum_k c^{(k)}=0$. For the case at hand $(N_1, N_2, N_3, N_4) = (1,1,1,N)$.

\section{The superpotential} \label{appsup}
the F-term potential in a supersymmetric theory has the following structure
\begin{align}
V &= \sum_\alpha \left| \frac{\partial W}{\partial \varphi_\alpha} \right|^2 = \frac{\partial W}{\partial \varphi_\alpha}  \frac{\partial \overline{W}}{\partial \overline{\varphi_\alpha}} 
\end{align}
where $\varphi_\alpha$ stands for various chiral multiplets (or complex scalars therein) in the theory. 

The superpotential in the present case is given by\footnote{We have dropped an overall factor of $\sqrt{2}$, which is  irrelevant for the analysing the minima.} \cite{Chowdhury:2015gbk}
\begin{align}
\nn
W &= W_1 + W_2 + W_3 + W_4 \, , \\ 
\nn
W_1 &= \Phi^{(12)} Z^{12} Z^{21} + \Phi^{(23)} Z^{32} Z^{23} + \Phi^{(31)} Z^{31} Z^{13} - \Tr{} \Phi^{4}_1 Z^{41} Z^{14} - \Tr{} \Phi^{4}_2 Z^{42} Z^{24} - \Tr{} \Phi^{4}_3 Z^{43} Z^{34} \\
\nn
W_2 &= Z^{12} Z^{23} Z^{31} +  Z^{13} Z^{32} Z^{21} + Z^{12} Z^{24}. Z^{41}  +  Z^{21} Z^{14}. Z^{42}  \\
\nn
&-  Z^{13} Z^{34}. Z^{41}  +  Z^{31} Z^{14} . Z^{43}  +  Z^{23} Z^{34} . Z^{42}  + Z^{32} Z^{24} . Z^{41} \\
\nn
W_3 &= c^{12} \Phi^{(12)}_3 + c^{23} \Phi^{(23)}_1 + c^{31} \Phi^{(31)}_2 - c^{14} \Tr{} \Phi_1^{4} - c^{24} \Tr{} \Phi_2^{4} - c^{34} \Tr{} \Phi_3^{4} \\
W_4 &= - \Tr{} \left( \Phi^{4}_1 \Phi^{4}_2 \Phi^{4}_3 -   \Phi^{4}_1 \Phi^{4}_3 \Phi^{4}_2 \right) \, , 
\end{align}
where $ \Phi^{(12)},  \Phi^{(23)},  \Phi^{(31)}$ are as defined in \refb{phisurvive}.
\section{Gauge fixing}\label{appgauuge}
The system has a $\mathbb{C}^\ast  \times \mathbb{C}^\ast \times \times GL(N, \mathbb{C})$ gauge symmetry.
First and second $\mathbb{C}^\ast$ gauge symmetries can respectively be fixed by fixing $Z^{12}, Z^{23}$ to some complex numbers. 

$GL(N, \mathbb{C})$ can be fixed in many ways. We choose to fix one of $Z^{41}, Z^{42}, Z^{43}, Z^{14}, Z^{24}, Z^{34}$ to a randomly chosen vector $Z^{fix}$ and one of $\Phi^{4}_1, \Phi^{4}_2, \Phi^{4}_3$ to a $N \times N$ matrix $\Phi_{fix}$ with all but $N$ components fixed.
Ability of taking any configuration of say $Z^{41}, \Phi^4_1 $ to this configuration depends on the existence of a unique $M \in GL(N, \mathbb{C})$ satisfying
\begin{align}
M Z^{41} &= Z^{fix} \quad , \quad M \Phi^4_1 M^{-1} = \Phi_{fix} \, \Rightarrow \, M \Phi^4_1 = \Phi_{fix} M \, .
\end{align}
These being linear equations in entries of $M$, will generally admit unique solution.

Simplest choice of $\Phi_{fix}$ would be a diagonal matrix. We point out a more general choice, which is to fix first $N-1$ rows of $\Phi_{fix}$ to randomly chosen numbers. For different choices of these random numbers, different minima might be more accessible.

As an explicit example, consider $N=3$. In  this case, 
\begin{align}
\Phi_{fix} &=
\begin{pmatrix}
r_{11} & r_{12} & r_{13} \\
r_{21} & r_{22} & r_{23} \\
v_1 & v_2 & v_3
\end{pmatrix} \, ,
\end{align}
where $r_{ij}$-s are randomly chosen complex numbers (to be held fixed in the course of SGD) and $v_i$-s are dynamic variables (to be varied in the course of SGD).

\section{Simplification of the potential landscape} \label{appsimple}
The variables $\Phi^{(12)} , \Phi^{(13)} , \Phi^{(23)}$ can be  eliminated by solving $G^{ij}=0 , \, i \neq j, \, i,j=1,2,3$ \refb{Fterm}, \refb{Feqn}. 
Each of $\Phi^{(12)} , \Phi^{(13)} , \Phi^{(23)} $ are fixed by two equations: 
\begin{align}
\nn
\Phi^{(12)} &= - \frac{Z^{23} Z^{31} + Z^{24} Z^{41}}{Z^{21}} = - \frac{Z^{13} Z^{32} + Z^{14} Z^{42}}{Z^{12}} \, , \,\\
\nn
\Phi^{(23)} &= - \frac{Z^{31} Z^{12} + Z^{34} Z^{42}}{Z^{32}} = - \frac{Z^{21} Z^{13} + Z^{24} Z^{43}}{Z^{23}} \, , \, \\
\Phi^{(31)} &= - \frac{Z^{32} Z^{21} - Z^{34} Z^{41}}{Z^{31}} = - \frac{Z^{12} Z^{23} + Z^{14} Z^{43}}{Z^{13}}  \, , \label{phifix}
\end{align}
giving 3 consistency conditions:
\begin{align}
E^{12} &:= Z^{12} Z^{23} Z^{31} + Z^{12} Z^{24} Z^{41} - Z^{21} Z^{13} Z^{32} - Z^{21} Z^{14} Z^{42} =0 ,  \\
\nonumber
E^{31} &:= Z^{13} Z^{32} Z^{21} - Z^{13} Z^{34} Z^{41} - Z^{31} Z^{12} Z^{23} - Z^{31} Z^{14} Z^{43} =0 ,  \\
\nonumber
E^{23} &:= Z^{23} Z^{31} Z^{12} + Z^{23} Z^{34} Z^{42} - Z^{32} Z^{21} Z^{13} - Z^{32} Z^{24} Z^{43} =0 \, . 
\end{align}
We  also note $F^{ij} = F^{ji}, \, i \neq j , \, i,j=1,2,3$, and $F^{i4} = \Tr{} F^{4i}, \, i=1,2,3$. So solving \refb{Feqn} is same as eliminating  $\Phi^{(12)} , \Phi^{(13)} , \Phi^{(23)}$ and solving  
\begin{align}
\nn
E^{12} &= 0 \, , \, E^{23} = 0 \, , \, E^{31} = 0 \, , \\
\nn
F^{12} &= 0 \, , \, F^{23} = 0 \, , \, F^{31} = 0 \, , \, F^{41} = 0 \, , \, F^{42} = 0 \, , \, F^{43} = 0 \, ,\\
G^{41} &= 0 \, , \, G^{42} = 0 \, , \, G^{43} = 0 \, , \, G^{14} = 0 \, , \, G^{24} = 0 \, , \, G^{34} = 0 \, \, ,  \label{effeqn}
\end{align}
which in turn is same as finding global minima of the effective potential
\begin{align}
\nn
V_{eff} &= |F^{12}|^2 +  |F^{23}|^2 +  |F^{31}|^2 + \Tr{} \left[ \left( F^{41} \right)^\dagger  F^{41} \right]   + \Tr{} \left[ \left( F^{42} \right)^\dagger  F^{42} \right]
+ \Tr{} \left[ \left( F^{43} \right)^\dagger  F^{43} \right] \\
\nn
&+  \left( G^{41} \right)^\dagger  G^{41} +  \left( G^{42} \right)^\dagger  G^{42} +  \left( G^{43} \right)^\dagger  G^{43}  
+  G^{14}  \left( G^{14} \right)^\dagger +  G^{24}  \left( G^{24} \right)^\dagger +  G^{34}  \left( G^{34} \right)^\dagger \\
&+  |E^{12}|^2 +  |E^{23}|^2 +  |E^{31}|^2 \, . \label{Vfsimple}
\end{align}

\section{Further simplification for $N=1$} \label{appN1}
For $N=1$, all the variables are Abelian and the equations \refb{Feqn} simplify considerably.
In particular, all $\Phi$ -s can be fixed in terms of $Z$-s and rest of the F-term equations reduce to the following 9 equations:
\begin{align}
\nn
Z^{ij} Z^{ji} &= -c_{ij} \, \quad 1 \leq i < j \leq 4 \, , \\
\nn
Z^{23} Z^{31} Z^{12} +  Z^{23} Z^{34} Z^{42} &= Z^{32} Z^{21} Z^{13} + Z^{32} Z^{24} Z^{43} \, , \\
\nn
Z^{24} Z^{41} Z^{12} +  Z^{24} Z^{43} Z^{32} &= Z^{42} Z^{21} Z^{14} + Z^{42} Z^{23} Z^{34} \, , \\
Z^{34} Z^{41} Z^{13} +  Z^{34} Z^{42} Z^{23} &= - Z^{43} Z^{31} Z^{14} + Z^{43} Z^{32} Z^{24} \, . \label{9eqn}
\end{align}
Define
\begin{align}
u_7 &= Z^{12} Z^{24} Z^{41} \, ,~u_8 = Z^{13} Z^{34} Z^{41} \, ,~u_9 = Z^{23} Z^{34} Z^{42} \, .
\end{align}
Using first 6 equations  of \refb{9eqn}, we have
\begin{align}
\nn
Z^{12} Z^{23} Z^{31} &= \frac{u_7}{Z^{24} Z^{41}} \frac{u_9}{Z^{34} Z^{42}} \frac{ (-c_{13}) }{Z^{13}} = \frac{c_{13} u_7 u_9}{ c_{24} u_8} \, , \\
\nn
Z^{32} Z^{21} Z^{13} &= \frac{(-c_{23})}{Z^{23}} \frac{(-c_{12})}{Z^{12}} \frac{u_8}{Z^{34} Z^{41}} = - \frac{c_{23} c_{12} c_{24} u_8}{u_7 u_9} \, , \\
Z^{32} Z^{24} Z^{43} &= - \frac{c_{23} c_{24} c_{34}}{u_9} \, , 
Z^{42} Z^{21} Z^{14} = - \frac{c_{24} c_{12} c_{14}}{u_7} \, , 
Z^{43} Z^{31} Z^{14} = - \frac{c_{34} c_{13} c_{14} }{u_8}  \, .
\end{align}
Using this,  last 3 equations  of \refb{9eqn} become
\begin{align}
\nonumber
c_{13} u_7^2 u_9^2 + c_{24} u_7 u_8 u_9^2 +  c_{23} c_{12} c_{24}^2 u_8^2 + c_{23} c_{24}^2 c_{34}u_7 u_8 &= 0 \, , \\
\nonumber
u_7^2 u_9 - u_7 u_9^2 -  c_{23} c_{24} c_{34} u_7 + c_{12} c_{14} c_{24} u_9  &= 0 \, , \\
u_8^2 u_9 + u_8 u_9^2 - c_{34} c_{13} c_{14} u_9 + c_{23} c_{24} c_{34} u_8 &= 0 \, .
\end{align}
Solutions with non-vanishing $u_7,u_8,u_9$ are the physical ones. It can easily be checked (e.g. in Mathematica) that there are 12 physical solutions.
\section{Derivation  of $p(3,n)$} \label{appp3n}
To start with we recall  \refb{pKn}, which entails 
\begin{align}
p_{ansatz} (3,n) &:= 1 - \frac{1}{3^n} \sum_{a=1}^{2} (-1)^{a-1} ~ {{3} \choose {a}} \,  (3-a)^n 
= 1 - \frac{ 2^n - 1}{3^{n-1}} 
 \, , \label{p3n}
\end{align}

Let $m_S$ denote the number of sequences with entries drawn from the set $S \subset \{1,2,3\}$. Thus,
\begin{align}
m_{\{12\}} = m_{\{13\}} = m_{\{23\}} = 2^n,~m_{\{1\}}=m_{\{2\}}=m_{\{3\}}=1 \, .
\end{align}
Note $m_{\{12\}}$ is the number of sequences without 3, $m_{\{13\}}$ is the number of sequences without 2, $m_{\{23\}}$ is the number of sequences without 1. Thus, up to overcounting, number of sequences without at least one of $1,2,3$ is $m_{\{12\}} + m_{\{13\}} + m_{\{23\}}$.
To take care of overcountings, we note the sequence $1^n$ is counted in both $m_{\{12\}}$ and $m_{\{13\}}$. Similar statements hold for sequences $2^n$ and $3^n$. With overcountings taken care of, the number of sequences without at least one of $1,2,3$ is
\begin{align}
\nonumber
m_{\{12\}} + m_{\{13\}} + m_{\{23\}} - m_{\{1\}} - m_{\{2\}} - m_{\{3\}} &= 3 (2^n - 1) \\
\Rightarrow p(3,n) &= 1 - \frac{3 (2^n -1)}{3^n} = 1 - \frac{2^n-1}{3^{n-1}} \, ,
\end{align}
which precisely matches \refb{p3n}.
Note $p(3,1)=p(3,2)=0$, as expected.
\section{Excluding already found solutions} \label{appexclude}
We consider a simple function $f$ with multiple degenerate minima and try to  modify the function to another function $\tilde{f}$, such that one of the minima of $f$ is not a minima of $\tilde{f}$, but others continue to be so, at least approximately. This can be achieved simply by adding a localized hump around the minimum to be excluded.

\begin{figure}[H]
\begin{center}
\includegraphics[scale=.8]{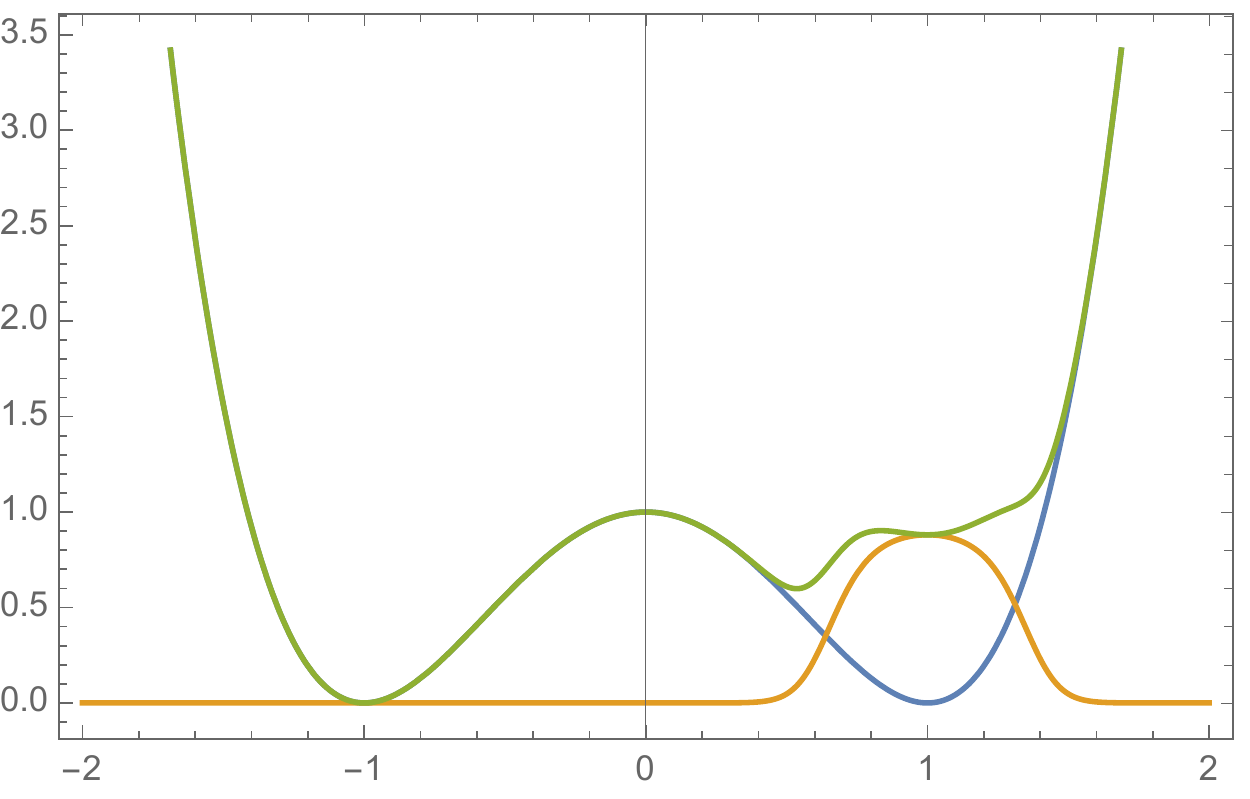}
\caption{The blue graph is the original potential, the orange graph is the hump added and the green graph is the resultant modified potential. The hump function is $\frac{1 - \tanh \left( 10 (x-1)^2 - 0.1 \right)}{2}$.}\label{exclusive}
\end{center}
\end{figure}

A simple choice for the hump function around a point $x_0$ would be 
\begin{align}
t(x_0) &:= \frac{1 - \tanh \left( \lambda (x-x_0)^2 - a \right)}{2} \, , 
\end{align}
where the parameter $a$ controls  the height  of the hump and $\lambda$ its width. If a function $f(x)$ has a minima at $x_0$, then we can take $\tilde{f}(x) = f(x) + t(x_0)$ to be the modified function. As an example, in Fig. \ref{exclusive}, we plot the function $f(x) = (x^2-1)^2, \, t(1), \, \tilde{f}(x) = f(x) + t(x_0)$. As can  be seen in Fig. \ref{exclusive}, the modified function (the green plot) and the original function (the blue plot) essentially coincide except in a small neighbourhood of the point $x=1$. Whereas the global minima at $x=1$ is avoided, new local minima shows up.

\bibliography{mlbh} 

\providecommand{\href}[2]{#2}\begingroup\raggedright\begin{thebibliography}{100}

\bibitem{NIPS2012_c399862d}
A.~Krizhevsky, I.~Sutskever and G.~E. Hinton, \emph{Imagenet classification
  with deep convolutional neural networks},  in \emph{Advances in Neural
  Information Processing Systems} (F.~Pereira, C.~Burges, L.~Bottou and
  K.~Weinberger, eds.), vol.~25, Curran Associates, Inc., 2012.

\bibitem{5740583}
G.~E. Dahl, D.~Yu, L.~Deng and A.~Acero, \emph{Context-dependent pre-trained
  deep neural networks for large-vocabulary speech recognition},
  \href{http://dx.doi.org/10.1109/TASL.2011.2134090}{\emph{IEEE Transactions on
  Audio, Speech, and Language Processing} {\bf 20} (2012) 30--42}.

\bibitem{Manning}
C.~Manning, \emph{Computational linguistics and deep learning},
  \href{http://dx.doi.org/10.1162/COLI_a_00239}{\emph{Computational
  Linguistics} {\bf 41} (09, 2015) 699--705}.

\bibitem{He:2017aed}
Y.-H. He, \emph{{Deep-Learning the Landscape}},
  \href{https://arxiv.org/abs/1706.02714}{{\tt 1706.02714}}.

\bibitem{He:2017set}
Y.-H. He, \emph{{Machine-learning the string landscape}},
  \href{http://dx.doi.org/10.1016/j.physletb.2017.10.024}{\emph{Phys. Lett. B}
  {\bf 774} (2017) 564--568}.

\bibitem{Ruehle:2017mzq}
F.~Ruehle, \emph{{Evolving neural networks with genetic algorithms to study the
  String Landscape}},
  \href{http://dx.doi.org/10.1007/JHEP08(2017)038}{\emph{JHEP} {\bf 08} (2017)
  038}, [\href{https://arxiv.org/abs/1706.07024}{{\tt 1706.07024}}].

\bibitem{Carifio:2017bov}
J.~Carifio, J.~Halverson, D.~Krioukov and B.~D. Nelson, \emph{{Machine Learning
  in the String Landscape}},
  \href{http://dx.doi.org/10.1007/JHEP09(2017)157}{\emph{JHEP} {\bf 09} (2017)
  157}, [\href{https://arxiv.org/abs/1707.00655}{{\tt 1707.00655}}].

\bibitem{Mutter:2018sra}
A.~M\"utter, E.~Parr and P.~K.~S. Vaudrevange, \emph{{Deep learning in the
  heterotic orbifold landscape}},
  \href{http://dx.doi.org/10.1016/j.nuclphysb.2019.01.013}{\emph{Nucl. Phys. B}
  {\bf 940} (2019) 113--129}, [\href{https://arxiv.org/abs/1811.05993}{{\tt
  1811.05993}}].

\bibitem{He:2021eiu}
Y.-H. He, S.~Lal and M.~Z. Zaz, \emph{{The World in a Grain of Sand: Condensing
  the String Vacuum Degeneracy}},  \href{https://arxiv.org/abs/2111.04761}{{\tt
  2111.04761}}.

\bibitem{Bull:2018uow}
K.~Bull, Y.-H. He, V.~Jejjala and C.~Mishra, \emph{{Machine Learning CICY
  Threefolds}},
  \href{http://dx.doi.org/10.1016/j.physletb.2018.08.008}{\emph{Phys. Lett. B}
  {\bf 785} (2018) 65--72}, [\href{https://arxiv.org/abs/1806.03121}{{\tt
  1806.03121}}].

\bibitem{Bull:2019cij}
K.~Bull, Y.-H. He, V.~Jejjala and C.~Mishra, \emph{{Getting CICY High}},
  \href{http://dx.doi.org/10.1016/j.physletb.2019.06.067}{\emph{Phys. Lett. B}
  {\bf 795} (2019) 700--706}, [\href{https://arxiv.org/abs/1903.03113}{{\tt
  1903.03113}}].

\bibitem{Jejjala:2020wcc}
V.~Jejjala, D.~K. Mayorga~Pena and C.~Mishra, \emph{{Neural network
  approximations for Calabi-Yau metrics}},
  \href{http://dx.doi.org/10.1007/JHEP08(2022)105}{\emph{JHEP} {\bf 08} (2022)
  105}, [\href{https://arxiv.org/abs/2012.15821}{{\tt 2012.15821}}].

\bibitem{Berglund:2022gvm}
P.~Berglund, G.~Butbaia, T.~H\"ubsch, V.~Jejjala, D.~Mayorga Pe\~na, C.~Mishra
  et~al., \emph{{Machine Learned Calabi-Yau Metrics and Curvature}},
  \href{https://arxiv.org/abs/2211.09801}{{\tt 2211.09801}}.

\bibitem{Erbin:2020srm}
H.~Erbin and R.~Finotello, \emph{{Inception neural network for complete
  intersection Calabi\textendash{}Yau 3-folds}},
  \href{http://dx.doi.org/10.1088/2632-2153/abda61}{\emph{Mach. Learn. Sci.
  Tech.} {\bf 2} (2021) 02LT03}, [\href{https://arxiv.org/abs/2007.13379}{{\tt
  2007.13379}}].

\bibitem{Erbin:2020tks}
H.~Erbin and R.~Finotello, \emph{{Machine learning for complete intersection
  Calabi-Yau manifolds: a methodological study}},
  \href{http://dx.doi.org/10.1103/PhysRevD.103.126014}{\emph{Phys. Rev. D} {\bf
  103} (2021) 126014}, [\href{https://arxiv.org/abs/2007.15706}{{\tt
  2007.15706}}].

\bibitem{He:2020lbz}
Y.-H. He and A.~Lukas, \emph{{Machine Learning Calabi-Yau Four-folds}},
  \href{http://dx.doi.org/10.1016/j.physletb.2021.136139}{\emph{Phys. Lett. B}
  {\bf 815} (2021) 136139}, [\href{https://arxiv.org/abs/2009.02544}{{\tt
  2009.02544}}].

\bibitem{Anderson:2020hux}
L.~B. Anderson, M.~Gerdes, J.~Gray, S.~Krippendorf, N.~Raghuram and F.~Ruehle,
  \emph{{Moduli-dependent Calabi-Yau and SU(3)-structure metrics from Machine
  Learning}}, \href{http://dx.doi.org/10.1007/JHEP05(2021)013}{\emph{JHEP} {\bf
  05} (2021) 013}, [\href{https://arxiv.org/abs/2012.04656}{{\tt 2012.04656}}].

\bibitem{Berman:2021mcw}
D.~S. Berman, Y.-H. He and E.~Hirst, \emph{{Machine learning Calabi-Yau
  hypersurfaces}},
  \href{http://dx.doi.org/10.1103/PhysRevD.105.066002}{\emph{Phys. Rev. D} {\bf
  105} (2022) 066002}, [\href{https://arxiv.org/abs/2112.06350}{{\tt
  2112.06350}}].

\bibitem{Erbin:2021hmx}
H.~Erbin, R.~Finotello, R.~Schneider and M.~Tamaazousti, \emph{{Deep multi-task
  mining Calabi\textendash{}Yau four-folds}},
  \href{http://dx.doi.org/10.1088/2632-2153/ac37f7}{\emph{Mach. Learn. Sci.
  Tech.} {\bf 3} (2022) 015006}, [\href{https://arxiv.org/abs/2108.02221}{{\tt
  2108.02221}}].

\bibitem{Craven:2020bdz}
J.~Craven, V.~Jejjala and A.~Kar, \emph{{Disentangling a deep learned volume
  formula}}, \href{http://dx.doi.org/10.1007/JHEP06(2021)040}{\emph{JHEP} {\bf
  06} (2021) 040}, [\href{https://arxiv.org/abs/2012.03955}{{\tt 2012.03955}}].

\bibitem{Craven:2021ckk}
J.~Craven, M.~Hughes, V.~Jejjala and A.~Kar, \emph{{Learning knot invariants
  across dimensions}},
  \href{http://dx.doi.org/10.21468/SciPostPhys.14.2.021}{\emph{SciPost Phys.}
  {\bf 14} (2023) 021}, [\href{https://arxiv.org/abs/2112.00016}{{\tt
  2112.00016}}].

\bibitem{Craven:2022cxe}
J.~Craven, M.~Hughes, V.~Jejjala and A.~Kar, \emph{{Illuminating new and known
  relations between knot invariants}},
  \href{https://arxiv.org/abs/2211.01404}{{\tt 2211.01404}}.

\bibitem{Gukov:2020qaj}
S.~Gukov, J.~Halverson, F.~Ruehle and P.~Su\l{}kowski, \emph{{Learning to
  Unknot}}, \href{http://dx.doi.org/10.1088/2632-2153/abe91f}{\emph{Mach.
  Learn. Sci. Tech.} {\bf 2} (2021) 025035},
  [\href{https://arxiv.org/abs/2010.16263}{{\tt 2010.16263}}].

\bibitem{Hashimoto:2018ftp}
K.~Hashimoto, S.~Sugishita, A.~Tanaka and A.~Tomiya, \emph{{Deep learning and
  the AdS/CFT correspondence}},
  \href{http://dx.doi.org/10.1103/PhysRevD.98.046019}{\emph{Phys. Rev. D} {\bf
  98} (2018) 046019}, [\href{https://arxiv.org/abs/1802.08313}{{\tt
  1802.08313}}].

\bibitem{Hashimoto:2018bnb}
K.~Hashimoto, S.~Sugishita, A.~Tanaka and A.~Tomiya, \emph{{Deep Learning and
  Holographic QCD}},
  \href{http://dx.doi.org/10.1103/PhysRevD.98.106014}{\emph{Phys. Rev. D} {\bf
  98} (2018) 106014}, [\href{https://arxiv.org/abs/1809.10536}{{\tt
  1809.10536}}].

\bibitem{Tan:2019czc}
J.~Tan and C.-B. Chen, \emph{{Deep learning the holographic black hole with
  charge}}, \href{http://dx.doi.org/10.1142/S0218271819501530}{\emph{Int. J.
  Mod. Phys. D} {\bf 28} (2019) 1950153},
  [\href{https://arxiv.org/abs/1908.01470}{{\tt 1908.01470}}].

\bibitem{Akutagawa:2020yeo}
T.~Akutagawa, K.~Hashimoto and T.~Sumimoto, \emph{{Deep Learning and AdS/QCD}},
  \href{http://dx.doi.org/10.1103/PhysRevD.102.026020}{\emph{Phys. Rev. D} {\bf
  102} (2020) 026020}, [\href{https://arxiv.org/abs/2005.02636}{{\tt
  2005.02636}}].

\bibitem{Yan:2020wcd}
Y.-K. Yan, S.-F. Wu, X.-H. Ge and Y.~Tian, \emph{{Deep learning black hole
  metrics from shear viscosity}},
  \href{http://dx.doi.org/10.1103/PhysRevD.102.101902}{\emph{Phys. Rev. D} {\bf
  102} (4, 2020) 101902}, [\href{https://arxiv.org/abs/2004.12112}{{\tt
  2004.12112}}].

\bibitem{Chen:2020dxg}
H.-Y. Chen, Y.-H. He, S.~Lal and M.~Z. Zaz, \emph{{Machine Learning Etudes in
  Conformal Field Theories}},  \href{https://arxiv.org/abs/2006.16114}{{\tt
  2006.16114}}.

\bibitem{Basu:2022qaf}
P.~Basu, J.~Bhattacharya, D.~P.~S. Jakka, C.~Mosomane and V.~Shukla,
  \emph{{Machine learning of Ising criticality with spin-shuffling}},
  \href{https://arxiv.org/abs/2203.04012}{{\tt 2203.04012}}.

\bibitem{Kuo:2021lvu}
E.-J. Kuo, A.~Seif, R.~Lundgren, S.~Whitsitt and M.~Hafezi, \emph{{Decoding
  conformal field theories: From supervised to unsupervised learning}},
  \href{http://dx.doi.org/10.1103/PhysRevResearch.4.043031}{\emph{Phys. Rev.
  Res.} {\bf 4} (2022) 043031}, [\href{https://arxiv.org/abs/2106.13485}{{\tt
  2106.13485}}].

\bibitem{Kantor:2021kbx}
G.~K\'antor, V.~Niarchos and C.~Papageorgakis, \emph{{Solving Conformal Field
  Theories with Artificial Intelligence}},
  \href{http://dx.doi.org/10.1103/PhysRevLett.128.041601}{\emph{Phys. Rev.
  Lett.} {\bf 128} (2022) 041601},
  [\href{https://arxiv.org/abs/2108.08859}{{\tt 2108.08859}}].

\bibitem{Kantor:2021jpz}
G.~K\'antor, V.~Niarchos and C.~Papageorgakis, \emph{{Conformal bootstrap with
  reinforcement learning}},
  \href{http://dx.doi.org/10.1103/PhysRevD.105.025018}{\emph{Phys. Rev. D} {\bf
  105} (2022) 025018}, [\href{https://arxiv.org/abs/2108.09330}{{\tt
  2108.09330}}].

\bibitem{Kantor:2022epi}
G.~K\'antor, V.~Niarchos, C.~Papageorgakis and P.~Richmond, \emph{{6D (2,0)
  bootstrap with the soft-actor-critic algorithm}},
  \href{http://dx.doi.org/10.1103/PhysRevD.107.025005}{\emph{Phys. Rev. D} {\bf
  107} (2023) 025005}, [\href{https://arxiv.org/abs/2209.02801}{{\tt
  2209.02801}}].

\bibitem{Chen:2020jjw}
H.-Y. Chen, Y.-H. He, S.~Lal and S.~Majumder, \emph{{Machine learning Lie
  structures \& applications to physics}},
  \href{http://dx.doi.org/10.1016/j.physletb.2021.136297}{\emph{Phys. Lett. B}
  {\bf 817} (2021) 136297}, [\href{https://arxiv.org/abs/2011.00871}{{\tt
  2011.00871}}].

\bibitem{Lal:2022otf}
S.~Lal, \emph{{Machine Learning Symmetry}},  in \emph{{Nankai Symposium on
  Mathematical Dialogues}: {In celebration of S.S.Chern's 110th anniversary}},
  1, 2022.
\newblock \href{https://arxiv.org/abs/2201.09345}{{\tt 2201.09345}}.

\bibitem{Koch:2019fxy}
E.~d.~M. Koch, R.~de~Mello~Koch and L.~Cheng, \emph{{Is Deep Learning a
  Renormalization Group Flow?}},  \href{https://arxiv.org/abs/1906.05212}{{\tt
  1906.05212}}.

\bibitem{Halverson:2020trp}
J.~Halverson, A.~Maiti and K.~Stoner, \emph{{Neural Networks and Quantum Field
  Theory}}, \href{http://dx.doi.org/10.1088/2632-2153/abeca3}{\emph{Mach.
  Learn. Sci. Tech.} {\bf 2} (2021) 035002},
  [\href{https://arxiv.org/abs/2008.08601}{{\tt 2008.08601}}].

\bibitem{Maiti:2021fpy}
A.~Maiti, K.~Stoner and J.~Halverson, \emph{{Symmetry-via-Duality: Invariant
  Neural Network Densities from Parameter-Space Correlators}},
  \href{https://arxiv.org/abs/2106.00694}{{\tt 2106.00694}}.

\bibitem{Halverson:2021aot}
J.~Halverson, \emph{{Building Quantum Field Theories Out of Neurons}},
  \href{https://arxiv.org/abs/2112.04527}{{\tt 2112.04527}}.

\bibitem{Erbin:2021kqf}
H.~Erbin, V.~Lahoche and D.~O. Samary, \emph{{Non-perturbative renormalization
  for the neural network-QFT correspondence}},
  \href{http://dx.doi.org/10.1088/2632-2153/ac4f69}{\emph{Mach. Learn. Sci.
  Tech.} {\bf 3} (2022) 015027}, [\href{https://arxiv.org/abs/2108.01403}{{\tt
  2108.01403}}].

\bibitem{Grosvenor:2021eol}
K.~T. Grosvenor and R.~Jefferson, \emph{{The edge of chaos: quantum field
  theory and deep neural networks}},
  \href{http://dx.doi.org/10.21468/SciPostPhys.12.3.081}{\emph{SciPost Phys.}
  {\bf 12} (2022) 081}, [\href{https://arxiv.org/abs/2109.13247}{{\tt
  2109.13247}}].

\bibitem{Erbin:2022lls}
H.~Erbin, V.~Lahoche and D.~O. Samary, \emph{{Renormalization in the neural
  network-quantum field theory correspondence}},  12, 2022.
\newblock \href{https://arxiv.org/abs/2212.11811}{{\tt 2212.11811}}.

\bibitem{Banta:2023kqe}
I.~Banta, T.~Cai, N.~Craig and Z.~Zhang, \emph{{Structures of Neural Network
  Effective Theories}},  \href{https://arxiv.org/abs/2305.02334}{{\tt
  2305.02334}}.

\bibitem{CaboBizet:2020cse}
N.~Cabo~Bizet, C.~Damian, O.~Loaiza-Brito, D.~K.~M. Pe\~na and J.~A. Monta\~nez
  Barrera, \emph{{Testing Swampland Conjectures with Machine Learning}},
  \href{http://dx.doi.org/10.1140/epjc/s10052-020-8332-9}{\emph{Eur. Phys. J.
  C} {\bf 80} (2020) 766}, [\href{https://arxiv.org/abs/2006.07290}{{\tt
  2006.07290}}].

\bibitem{Hashimoto:2019bih}
K.~Hashimoto, \emph{{AdS/CFT correspondence as a deep Boltzmann machine}},
  \href{http://dx.doi.org/10.1103/PhysRevD.99.106017}{\emph{Phys. Rev. D} {\bf
  99} (2019) 106017}, [\href{https://arxiv.org/abs/1903.04951}{{\tt
  1903.04951}}].

\bibitem{Betzler:2020rfg}
P.~Betzler and S.~Krippendorf, \emph{{Connecting Dualities and Machine
  Learning}}, \href{http://dx.doi.org/10.1002/prop.202000022}{\emph{Fortsch.
  Phys.} {\bf 68} (2020) 2000022},
  [\href{https://arxiv.org/abs/2002.05169}{{\tt 2002.05169}}].

\bibitem{Krippendorf:2020gny}
S.~Krippendorf and M.~Syvaeri, \emph{{Detecting Symmetries with Neural
  Networks}},  \href{https://arxiv.org/abs/2003.13679}{{\tt 2003.13679}}.

\bibitem{Bao:2020nbi}
J.~Bao, S.~Franco, Y.-H. He, E.~Hirst, G.~Musiker and Y.~Xiao, \emph{{Quiver
  Mutations, Seiberg Duality and Machine Learning}},
  \href{http://dx.doi.org/10.1103/PhysRevD.102.086013}{\emph{Phys. Rev. D} {\bf
  102} (2020) 086013}, [\href{https://arxiv.org/abs/2006.10783}{{\tt
  2006.10783}}].

\bibitem{Ruehle:2020jrk}
F.~Ruehle, \emph{{Data science applications to string theory}},
  \href{http://dx.doi.org/10.1016/j.physrep.2019.09.005}{\emph{Phys. Rept.}
  {\bf 839} (2020) 1--117}.

\bibitem{He:2023csq}
Y.-H. He, E.~Heyes and E.~Hirst, \emph{{Machine Learning in Physics and
  Geometry}},  \href{https://arxiv.org/abs/2303.12626}{{\tt 2303.12626}}.

\bibitem{bedolla2020machine}
E.~Bedolla, L.~C. Padierna and R.~Castaneda-Priego, \emph{Machine learning for
  condensed matter physics}, {\emph{Journal of Physics: Condensed Matter} {\bf
  33} (2020) 053001}.

\bibitem{samarakoon2021machine}
A.~M. Samarakoon and D.~A. Tennant, \emph{Machine learning for magnetic phase
  diagrams and inverse scattering problems}, {\emph{Journal of Physics:
  Condensed Matter} {\bf 34} (2021) 044002}.

\bibitem{carrasquilla2017machine}
J.~Carrasquilla and R.~G. Melko, \emph{Machine learning phases of matter},
  {\emph{Nature Physics} {\bf 13} (2017) 431--434}.

\bibitem{decelle2022introduction}
A.~Decelle, \emph{An introduction to machine learning: a perspective from
  statistical physics}, {\emph{Physica A: Statistical Mechanics and its
  Applications} (2022) 128154}.

\bibitem{RevModPhys.91.045002}
G.~Carleo, I.~Cirac, K.~Cranmer, L.~Daudet, M.~Schuld, N.~Tishby et~al.,
  \emph{Machine learning and the physical sciences},
  \href{http://dx.doi.org/10.1103/RevModPhys.91.045002}{\emph{Rev. Mod. Phys.}
  {\bf 91} (Dec, 2019) 045002}.

\bibitem{EonBottou1998OnlineLA}
L.~E. Bottou, \emph{Online learning and stochastic approximations},  1998.

\bibitem{10.1162/089976602760408035}
J.~Šíma, \emph{{Training a Single Sigmoidal Neuron Is Hard}},
  \href{http://dx.doi.org/10.1162/089976602760408035}{\emph{Neural Computation}
  {\bf 14} (11, 2002) 2709--2728},
  [\href{https://arxiv.org/abs/https://direct.mit.edu/neco/article-pdf/14/11/2709/815296/089976602760408035.pdf}{{\tt
  https://direct.mit.edu/neco/article-pdf/14/11/2709/815296/089976602760408035.pdf}}].

\bibitem{livni2014computational}
R.~Livni, S.~Shalev-Shwartz and O.~Shamir, \emph{On the computational
  efficiency of training neural networks},
  \href{https://arxiv.org/abs/1410.1141}{{\tt 1410.1141}}.

\bibitem{shalevshwartz2017failures}
S.~Shalev-Shwartz, O.~Shamir and S.~Shammah, \emph{Failures of gradient-based
  deep learning},  \href{https://arxiv.org/abs/1703.07950}{{\tt 1703.07950}}.

\bibitem{Murty1987SomeNP}
K.~G. Murty and S.~N. Kabadi, \emph{Some np-complete problems in quadratic and
  nonlinear programming}, {\emph{Mathematical Programming} {\bf 39} (1987)
  117--129}.

\bibitem{10.5555/2969735.2969792}
A.~Blum and R.~L. Rivest, \emph{Training a 3-node neural network is
  np-complete},  in \emph{Proceedings of the 1st International Conference on
  Neural Information Processing Systems}, NIPS'88, (Cambridge, MA, USA),
  p.~494–501, MIT Press, 1988.

\bibitem{freeman2017topology}
C.~D. Freeman and J.~Bruna, \emph{Topology and geometry of half-rectified
  network optimization},  \href{https://arxiv.org/abs/1611.01540}{{\tt
  1611.01540}}.

\bibitem{hoffer2018train}
E.~Hoffer, I.~Hubara and D.~Soudry, \emph{Train longer, generalize better:
  closing the generalization gap in large batch training of neural networks},
  \href{https://arxiv.org/abs/1705.08741}{{\tt 1705.08741}}.

\bibitem{soudry2016bad}
D.~Soudry and Y.~Carmon, \emph{No bad local minima: Data independent training
  error guarantees for multilayer neural networks},
  \href{https://arxiv.org/abs/1605.08361}{{\tt 1605.08361}}.

\bibitem{BALDI198953}
P.~Baldi and K.~Hornik, \emph{Neural networks and principal component analysis:
  Learning from examples without local minima},
  \href{http://dx.doi.org/https://doi.org/10.1016/0893-6080(89)90014-2}{\emph{Neural
  Networks} {\bf 2} (1989) 53--58}.

\bibitem{kawaguchi2016deep}
K.~Kawaguchi, \emph{Deep learning without poor local minima},
  \href{https://arxiv.org/abs/1605.07110}{{\tt 1605.07110}}.

\bibitem{nguyen2017loss}
Q.~Nguyen and M.~Hein, \emph{The loss surface of deep and wide neural
  networks},  \href{https://arxiv.org/abs/1704.08045}{{\tt 1704.08045}}.

\bibitem{107014}
M.~Gori and A.~Tesi, \emph{On the problem of local minima in backpropagation},
  \href{http://dx.doi.org/10.1109/34.107014}{\emph{IEEE Transactions on Pattern
  Analysis and Machine Intelligence} {\bf 14} (1992) 76--86}.

\bibitem{fra}
P.~Frasconi, M.~Gori and A.~Tesi, \emph{Successes and failures of
  backpropagation: A theoretical investigation}, .

\bibitem{410380}
X.-H. Yu and G.-A. Chen, \emph{On the local minima free condition of
  backpropagation learning},
  \href{http://dx.doi.org/10.1109/72.410380}{\emph{IEEE Transactions on Neural
  Networks} {\bf 6} (1995) 1300--1303}.

\bibitem{saxe2014exact}
A.~M. Saxe, J.~L. McClelland and S.~Ganguli, \emph{Exact solutions to the
  nonlinear dynamics of learning in deep linear neural networks},
  \href{https://arxiv.org/abs/1312.6120}{{\tt 1312.6120}}.

\bibitem{safran2018spurious}
I.~Safran and O.~Shamir, \emph{Spurious local minima are common in two-layer
  relu neural networks},  \href{https://arxiv.org/abs/1712.08968}{{\tt
  1712.08968}}.

\bibitem{yun2019small}
C.~Yun, S.~Sra and A.~Jadbabaie, \emph{Small nonlinearities in activation
  functions create bad local minima in neural networks},
  \href{https://arxiv.org/abs/1802.03487}{{\tt 1802.03487}}.

\bibitem{zou2018stochastic}
D.~Zou, Y.~Cao, D.~Zhou and Q.~Gu, \emph{Stochastic gradient descent optimizes
  over-parameterized deep relu networks},
  \href{https://arxiv.org/abs/1811.08888}{{\tt 1811.08888}}.

\bibitem{swirszcz2017local}
G.~Swirszcz, W.~M. Czarnecki and R.~Pascanu, \emph{Local minima in training of
  neural networks},  \href{https://arxiv.org/abs/1611.06310}{{\tt 1611.06310}}.

\bibitem{liu2021spurious}
B.~Liu, \emph{Spurious local minima are common for deep neural networks with
  piecewise linear activations},  \href{https://arxiv.org/abs/2102.13233}{{\tt
  2102.13233}}.

\bibitem{NIPS1995_3806734b}
P.~Auer, M.~Herbster and M.~K.~K. Warmuth, \emph{Exponentially many local
  minima for single neurons},  in \emph{Advances in Neural Information
  Processing Systems} (D.~Touretzky, M.~Mozer and M.~Hasselmo, eds.), vol.~8,
  MIT Press, 1995.

\bibitem{NIPS1996_a51fb975}
F.~Coetzee and V.~Stonick, \emph{488 solutions to the xor problem},  in
  \emph{Advances in Neural Information Processing Systems} (M.~Mozer, M.~Jordan
  and T.~Petsche, eds.), vol.~9, MIT Press, 1996.

\bibitem{choromanska2015loss}
A.~Choromanska, M.~Henaff, M.~Mathieu, G.~B. Arous and Y.~LeCun, \emph{The loss
  surfaces of multilayer networks},
  \href{https://arxiv.org/abs/1412.0233}{{\tt 1412.0233}}.

\bibitem{pmlr-v38-choromanska15}
A.~Choromanska, M.~Henaff, M.~Mathieu, G.~Ben~Arous and Y.~LeCun, \emph{{The
  Loss Surfaces of Multilayer Networks}},  in \emph{Proceedings of the
  Eighteenth International Conference on Artificial Intelligence and
  Statistics} (G.~Lebanon and S.~V.~N. Vishwanathan, eds.), vol.~38 of
  \emph{Proceedings of Machine Learning Research}, (San Diego, California,
  USA), pp.~192--204, PMLR, 09--12 May, 2015.

\bibitem{Ramu-2005}
C.~K. I.~W. Carl Edward~Rasmussen, \emph{Gaussian Processes for Machine
  Learning (Adaptive Computation and Machine Learning}.
\newblock MIT Press, 2005.

\bibitem{Bray_2007}
A.~J. Bray and D.~S. Dean, \emph{Statistics of critical points of gaussian
  fields on large-dimensional spaces},
  \href{http://dx.doi.org/10.1103/physrevlett.98.150201}{\emph{Physical Review
  Letters} {\bf 98} (apr, 2007) }.

\bibitem{fyodorov2007replica}
Y.~V. Fyodorov and I.~Williams, \emph{Replica symmetry breaking condition
  exposed by random matrix calculation of landscape complexity},
  \href{https://arxiv.org/abs/cond-mat/0702601}{{\tt cond-mat/0702601}}.

\bibitem{dauphin2014identifying}
Y.~Dauphin, R.~Pascanu, C.~Gulcehre, K.~Cho, S.~Ganguli and Y.~Bengio,
  \emph{Identifying and attacking the saddle point problem in high-dimensional
  non-convex optimization},  \href{https://arxiv.org/abs/1406.2572}{{\tt
  1406.2572}}.

\bibitem{pascanu2014saddle}
R.~Pascanu, Y.~N. Dauphin, S.~Ganguli and Y.~Bengio, \emph{On the saddle point
  problem for non-convex optimization},
  \href{https://arxiv.org/abs/1405.4604}{{\tt 1405.4604}}.

\bibitem{Goodfellow-et-al-2016}
I.~Goodfellow, Y.~Bengio and A.~Courville, \emph{Deep Learning}.
\newblock MIT Press, 2016.

\bibitem{PhysRevA.32.1007}
D.~J. Amit, H.~Gutfreund and H.~Sompolinsky, \emph{Spin-glass models of neural
  networks}, \href{http://dx.doi.org/10.1103/PhysRevA.32.1007}{\emph{Phys. Rev.
  A} {\bf 32} (Aug, 1985) 1007--1018}.

\bibitem{Nakanishi_1997}
K.~Nakanishi and H.~Takayama, \emph{Mean-field theory for a spin-glass model of
  neural networks: {TAP} free energy and the paramagnetic to spin-glass
  transition},
  \href{http://dx.doi.org/10.1088/0305-4470/30/23/009}{\emph{Journal of Physics
  A: Mathematical and General} {\bf 30} (dec, 1997) 8085--8094}.

\bibitem{pmlr-v40-Choromanska15}
A.~Choromanska, Y.~LeCun and G.~Ben~Arous, \emph{Open problem: The landscape of
  the loss surfaces of multilayer networks},  in \emph{Proceedings of The 28th
  Conference on Learning Theory} (P.~Grünwald, E.~Hazan and S.~Kale, eds.),
  vol.~40 of \emph{Proceedings of Machine Learning Research}, (Paris, France),
  pp.~1756--1760, PMLR, 03--06 Jul, 2015.

\bibitem{auffinger2011random}
A.~Auffinger, G.~B. Arous and J.~Cerny, \emph{Random matrices and complexity of
  spin glasses},  \href{https://arxiv.org/abs/1003.1129}{{\tt 1003.1129}}.

\bibitem{Auffinger_2013}
A.~Auffinger and G.~B. Arous, \emph{Complexity of random smooth functions on
  the high-dimensional sphere},
  \href{http://dx.doi.org/10.1214/13-aop862}{\emph{The Annals of Probability}
  {\bf 41} (nov, 2013) }.

\bibitem{Baity_Jesi_2019}
M.~Baity-Jesi, L.~Sagun, M.~Geiger, S.~Spigler, G.~B. Arous, C.~Cammarota
  et~al., \emph{Comparing dynamics: deep neural networks versus glassy
  systems}, \href{http://dx.doi.org/10.1088/1742-5468/ab3281}{\emph{Journal of
  Statistical Mechanics: Theory and Experiment} {\bf 2019} (dec, 2019) 124013}.

\bibitem{bouchaud1997equilibrium}
J.-P. Bouchaud, L.~F. Cugliandolo, J.~Kurchan and M.~Mezard, \emph{Out of
  equilibrium dynamics in spin-glasses and other glassy systems},
  \href{https://arxiv.org/abs/cond-mat/9702070}{{\tt cond-mat/9702070}}.

\bibitem{cugliandolo2002dynamics}
L.~F. Cugliandolo, \emph{Dynamics of glassy systems},
  \href{https://arxiv.org/abs/cond-mat/0210312}{{\tt cond-mat/0210312}}.

\bibitem{Berthier_2011}
L.~Berthier and G.~Biroli, \emph{Theoretical perspective on the glass
  transition and amorphous materials},
  \href{http://dx.doi.org/10.1103/revmodphys.83.587}{\emph{Reviews of Modern
  Physics} {\bf 83} (jun, 2011) 587--645}.

\bibitem{mehta2018loss}
D.~Mehta, T.~Chen, T.~Tang and J.~D. Hauenstein, \emph{The loss surface of deep
  linear networks viewed through the algebraic geometry lens},
  \href{https://arxiv.org/abs/1810.07716}{{\tt 1810.07716}}.

\bibitem{C7CP01108C}
A.~J. Ballard, R.~Das, S.~Martiniani, D.~Mehta, L.~Sagun, J.~D. Stevenson
  et~al., \emph{Energy landscapes for machine learning},
  \href{http://dx.doi.org/10.1039/C7CP01108C}{\emph{Phys. Chem. Chem. Phys.}
  {\bf 19} (2017) 12585--12603}.

\bibitem{Wales}
D.~J. Wales, \emph{Energy Landscapes:Applications to Clusters, Biomolecules and
  Glasses}.
\newblock Cambridge University Press, 2003.

\bibitem{doi:10.1146/annurev-statistics-032921-013738}
E.~Nalisnick, P.~Smyth and D.~Tran, \emph{A brief tour of deep learning from a
  statistical perspective},
  \href{http://dx.doi.org/10.1146/annurev-statistics-032921-013738}{\emph{Annual
  Review of Statistics and Its Application} {\bf 10} (2023) 219--246},
  [\href{https://arxiv.org/abs/https://doi.org/10.1146/annurev-statistics-032921-013738}{{\tt
  https://doi.org/10.1146/annurev-statistics-032921-013738}}].

\bibitem{doi:10.1146/annurev-conmatphys-031119-050745}
Y.~Bahri, J.~Kadmon, J.~Pennington, S.~S. Schoenholz, J.~Sohl-Dickstein and
  S.~Ganguli, \emph{Statistical mechanics of deep learning},
  \href{http://dx.doi.org/10.1146/annurev-conmatphys-031119-050745}{\emph{Annual
  Review of Condensed Matter Physics} {\bf 11} (2020) 501--528},
  [\href{https://arxiv.org/abs/https://doi.org/10.1146/annurev-conmatphys-031119-050745}{{\tt
  https://doi.org/10.1146/annurev-conmatphys-031119-050745}}].

\bibitem{Bekenstein1}
J.~W. Bekenstein, \emph{{Black Holes and Entropy}}, {\emph{Phys. Rev.} {\bf D7}
  (1973) 2333--2346}.

\bibitem{Bekenstein2}
J.~W. Bekenstein, \emph{{Generalized Second Law of Thermodynamics in Black-Hole
  Physics}}, {\emph{Phys. Rev.} {\bf D9} (1974) 3292--3300}.

\bibitem{BCH}
C.~B. Bardeen, J.M. and S.~W. Hawking, \emph{{The Four Laws of Black Hole
  Mechanics}}, {\emph{Commun. Math. Phys.} {\bf 31} (1973) 161--170}.

\bibitem{Hawking75}
S.~W. Hawking, \emph{{Particle Creation by Black Holes}}, {\emph{Commun. Math.
  Phys.} {\bf 43} (1975) 199--220}.

\bibitem{Strominger:1996sh}
A.~Strominger and C.~Vafa, \emph{{Microscopic origin of the Bekenstein-Hawking
  entropy}}, \href{http://dx.doi.org/10.1016/0370-2693(96)00345-0}{\emph{Phys.
  Lett. B} {\bf 379} (1996) 99--104},
  [\href{https://arxiv.org/abs/hep-th/9601029}{{\tt hep-th/9601029}}].

\bibitem{Maldacena:1997de}
J.~M. Maldacena, A.~Strominger and E.~Witten, \emph{{Black hole entropy in M
  theory}}, \href{http://dx.doi.org/10.1088/1126-6708/1997/12/002}{\emph{JHEP}
  {\bf 12} (1997) 002}, [\href{https://arxiv.org/abs/hep-th/9711053}{{\tt
  hep-th/9711053}}].

\bibitem{Shih:2005qf}
D.~Shih, A.~Strominger and X.~Yin, \emph{{Counting dyons in N=8 string
  theory}}, \href{http://dx.doi.org/10.1088/1126-6708/2006/06/037}{\emph{JHEP}
  {\bf 06} (2006) 037}, [\href{https://arxiv.org/abs/hep-th/0506151}{{\tt
  hep-th/0506151}}].

\bibitem{Chowdhury:2014yca}
A.~Chowdhury, R.~S. Garavuso, S.~Mondal and A.~Sen, \emph{{BPS State Counting
  in N=8 Supersymmetric String Theory for Pure D-brane Configurations}},
  \href{http://dx.doi.org/10.1007/JHEP10(2014)186}{\emph{JHEP} {\bf 10} (2014)
  186}, [\href{https://arxiv.org/abs/1405.0412}{{\tt 1405.0412}}].

\bibitem{Chowdhury:2015gbk}
A.~Chowdhury, R.~S. Garavuso, S.~Mondal and A.~Sen, \emph{{Do All BPS Black
  Hole Microstates Carry Zero Angular Momentum?}},
  \href{http://dx.doi.org/10.1007/JHEP04(2016)082}{\emph{JHEP} {\bf 04} (2016)
  082}, [\href{https://arxiv.org/abs/1511.06978}{{\tt 1511.06978}}].

\bibitem{Sen:2009gy}
A.~Sen, \emph{{Arithmetic of N=8 Black Holes}},
  \href{http://dx.doi.org/10.1007/JHEP02(2010)090}{\emph{JHEP} {\bf 02} (2010)
  090}, [\href{https://arxiv.org/abs/0908.0039}{{\tt 0908.0039}}].

\bibitem{PhysRevLett.26.1344}
S.~W. Hawking, \emph{Gravitational radiation from colliding black holes},
  \href{http://dx.doi.org/10.1103/PhysRevLett.26.1344}{\emph{Phys. Rev. Lett.}
  {\bf 26} (May, 1971) 1344--1346}.

\bibitem{Witten_morse}
E.~Witten, \emph{{Supersymmetry and Morse theory}}, {\emph{J. Differential.
  Geometry.} {\bf 17} (1982) 661--692}.

\bibitem{Schellekens_Warner}
A.~Schellekens and N.~P. Warner, \emph{{Anomalies, characters and strings}},
  {\emph{Phys. Lett..} {\bf B177} (1986) 317}.

\bibitem{Witten_elliptic}
E.~Witten, \emph{{Elliptic genera and quantum field theory}}, {\emph{Commun.
  Math. Phys.} {\bf 109} (1987) 525}.

\bibitem{Denef:2002ru}
F.~Denef, \emph{{Quantum quivers and Hall / hole halos}},
  \href{http://dx.doi.org/10.1088/1126-6708/2002/10/023}{\emph{JHEP} {\bf 10}
  (2002) 023}, [\href{https://arxiv.org/abs/hep-th/0206072}{{\tt
  hep-th/0206072}}].

\bibitem{Bena:2012hf}
I.~Bena, M.~Berkooz, J.~de~Boer, S.~El-Showk and D.~Van~den Bleeken,
  \emph{{Scaling BPS Solutions and pure-Higgs States}},
  \href{http://dx.doi.org/10.1007/JHEP11(2012)171}{\emph{JHEP} {\bf 11} (2012)
  171}, [\href{https://arxiv.org/abs/1205.5023}{{\tt 1205.5023}}].

\bibitem{Dabholkar:2010rm}
A.~Dabholkar, J.~Gomes, S.~Murthy and A.~Sen, \emph{{Supersymmetric Index from
  Black Hole Entropy}},
  \href{http://dx.doi.org/10.1007/JHEP04(2011)034}{\emph{JHEP} {\bf 04} (2011)
  034}, [\href{https://arxiv.org/abs/1009.3226}{{\tt 1009.3226}}].

\bibitem{Sen:2011ktd}
A.~Sen, \emph{{How Do Black Holes Predict the Sign of the Fourier Coefficients
  of Siegel Modular Forms?}},
  \href{http://dx.doi.org/10.1007/s10714-011-1175-9}{\emph{Gen. Rel. Grav.}
  {\bf 43} (2011) 2171--2183}, [\href{https://arxiv.org/abs/1008.4209}{{\tt
  1008.4209}}].

\bibitem{Bringmann:2012zr}
K.~Bringmann and S.~Murthy, \emph{{On the positivity of black hole degeneracies
  in string theory}},
  \href{http://dx.doi.org/10.4310/CNTP.2013.v7.n1.a2}{\emph{Commun. Num. Theor
  Phys.} {\bf 07} (2013) 15--56}, [\href{https://arxiv.org/abs/1208.3476}{{\tt
  1208.3476}}].

\bibitem{Chattopadhyaya:2021rdi}
A.~Chattopadhyaya, J.~Manschot and S.~Mondal, \emph{{Scaling black holes and
  modularity}}, \href{http://dx.doi.org/10.1007/JHEP03(2022)001}{\emph{JHEP}
  {\bf 03} (2022) 001}, [\href{https://arxiv.org/abs/2110.05504}{{\tt
  2110.05504}}].

\bibitem{Beaujard:2021fsk}
G.~Beaujard, S.~Mondal and B.~Pioline, \emph{{Multi-centered black holes,
  scaling solutions and pure-Higgs indices from localization}},
  \href{http://dx.doi.org/10.21468/SciPostPhys.11.2.023}{\emph{SciPost Phys.}
  {\bf 11} (2021) 023}, [\href{https://arxiv.org/abs/2103.03205}{{\tt
  2103.03205}}].

\bibitem{garipov2018loss}
T.~Garipov, P.~Izmailov, D.~Podoprikhin, D.~Vetrov and A.~G. Wilson, \emph{Loss
  surfaces, mode connectivity, and fast ensembling of dnns},
  \href{https://arxiv.org/abs/1802.10026}{{\tt 1802.10026}}.

\bibitem{raghavan2020sparsifying}
G.~Raghavan and M.~Thomson, \emph{Sparsifying networks by traversing
  geodesics},  \href{https://arxiv.org/abs/2012.09605}{{\tt 2012.09605}}.

\bibitem{NIPS1994_01882513}
S.~Hochreiter and J.~Schmidhuber, \emph{Simplifying neural nets by discovering
  flat minima},  in \emph{Advances in Neural Information Processing Systems}
  (G.~Tesauro, D.~Touretzky and T.~Leen, eds.), vol.~7, MIT Press, 1994.

\bibitem{chaudhari2017entropysgd}
P.~Chaudhari, A.~Choromanska, S.~Soatto, Y.~LeCun, C.~Baldassi, C.~Borgs
  et~al., \emph{Entropy-sgd: Biasing gradient descent into wide valleys},
  \href{https://arxiv.org/abs/1611.01838}{{\tt 1611.01838}}.

\bibitem{keskar2017largebatch}
N.~S. Keskar, D.~Mudigere, J.~Nocedal, M.~Smelyanskiy and P.~T.~P. Tang,
  \emph{On large-batch training for deep learning: Generalization gap and sharp
  minima},  \href{https://arxiv.org/abs/1609.04836}{{\tt 1609.04836}}.

\bibitem{Mathematica}
W.~R. Inc., ``Mathematica, {V}ersion 12.0.''

\bibitem{tensorflow2015-whitepaper}
M.~Abadi, A.~Agarwal, P.~Barham, E.~Brevdo, Z.~Chen, C.~Citro et~al.,
  \emph{{TensorFlow}: Large-scale machine learning on heterogeneous systems},
  2015.

\bibitem{Bachas:1996bp}
C.~Bachas and E.~Kiritsis, \emph{{F(4) terms in N=4 string vacua}},
  \href{http://dx.doi.org/10.1016/S0920-5632(97)00079-0}{\emph{Nucl. Phys. B
  Proc. Suppl.} {\bf 55} (1997) 194--199},
  [\href{https://arxiv.org/abs/hep-th/9611205}{{\tt hep-th/9611205}}].

\bibitem{Gregori:1997hi}
A.~Gregori, E.~Kiritsis, C.~Kounnas, N.~A. Obers, P.~M. Petropoulos and
  B.~Pioline, \emph{{R**2 corrections and nonperturbative dualities of N=4
  string ground states}},
  \href{http://dx.doi.org/10.1016/S0550-3213(97)00635-4}{\emph{Nucl. Phys. B}
  {\bf 510} (1998) 423--476}, [\href{https://arxiv.org/abs/hep-th/9708062}{{\tt
  hep-th/9708062}}].

\end{thebibliography}\endgroup
\bibliographystyle{JHEP}

\end{document}